\documentclass[journal,a4paper,10pt]{IEEEtran}
\usepackage{amsmath,amssymb,amsfonts} 
\usepackage{graphicx}                 %
\usepackage{url}                 
\usepackage{fontenc}
\usepackage{multirow} 
\usepackage{theorem}
\usepackage{array}

\usepackage{comment}
\usepackage{color,cite}
\usepackage{algorithmic}
\usepackage{algorithm}
\usepackage{epstopdf}
\usepackage{float}
\usepackage{lipsum}
\usepackage{esint}
\usepackage{mathrsfs}
\usepackage[english]{babel}
\usepackage[table,xcdraw]{xcolor}
\usepackage{booktabs}
\usepackage{multirow,makecell}
\usepackage{longtable}
\usepackage{caption}
\usepackage{subcaption}

\DeclareUnicodeCharacter{2212}{-}

\usepackage{relsize}

\usepackage{colortbl}

\definecolor{Gray}{gray}{0.75}
\definecolor{Blue}{rgb}{0 ,0.99,0.98}
\definecolor{LightCyan}{rgb}{0.88,1,1}
\newcolumntype{b}{>{\columncolor{Gray}}c}
\newcolumntype{a}{>{\columncolor{Blue}}c}

\usepackage{array}
\newcolumntype{P}[1]{>{\centering\arraybackslash}p{#1}}
\newcolumntype{M}[1]{>{\centering\arraybackslash}m{#1}}

\begin{document}

\title{Reconfigurable Intelligent Surfaces: \\ Interplay of Unit-Cell- and Surface-Level Design and Performance under Quantifiable Benchmarks}
\author{{Ammar Rafique, Naveed Ul Hassan,~\IEEEmembership{Senior Member,~IEEE}, Muhammad Zubair,~\IEEEmembership{Senior Member,~IEEE}, Ijaz Haider Naqvi, Muhammad Qasim Mehmood, Chau Yuen,~\IEEEmembership{Fellow,~IEEE}, Marco Di Renzo,~\IEEEmembership{Fellow,~IEEE}, and M\'erouane Debbah~\IEEEmembership{Fellow,~IEEE}}
\thanks{A. Rafique, M. Q. Mehmood and M. Zubair are with the Electrical Engineering Department of Information Technology University (ITU) of the Punjab, Lahore, Pakistan 54600. (Emails: phdee19002@itu.edu.pk, qasim.mehmood@itu.edu.pk, muhammad.zubair@itu.edu.pk).}
\thanks{N. U. Hassan and I. H. Naqvi are with the Department of Electrical Engineering at Lahore University of Management Sciences (LUMS), Lahore, Pakistan 54792. (Emails: naveed.hassan@lums.edu.pk, ijaznaqvi@lums.edu.pk).}
\thanks{C. Yuen is with the Engineering Product Development at the Singapore University of Technology and Design (SUTD), 8 Somapah Road, Singapore 487372. (Email: yuenchau@sutd.edu.sg).}
\thanks{M. Di Renzo is with Universit\'e Paris-Saclay, CNRS, CentraleSup\'elec, Laboratoire des Signaux et Systemes, 3 Rue Joliot-Curie, 91192 Gif-sur-Yvette, France. (marco.di-renzo@universite-paris-saclay.fr)}
\thanks{M\'erouane Debbah is with the Technology Innovation Institute, Masdar City 9639, Abu Dhabi, UAE, and also with the CentraleSup\'elec, University ParisSaclay, 91192 Gif-sur-Yvette, France (e-mail: merouane.debbah@tii.ae).}
\thanks{Correspondence may be sent to N. U. Hassan, M. Zubair and M. D. Renzo}

}

\maketitle

\begin{abstract}
The ability of reconfigurable intelligent surfaces (RIS) to produce complex radiation patterns in the far-field is determined by various factors, such as the unit-cell's size, shape, spatial arrangement, tuning mechanism, the communication and control circuitry's complexity, and the illuminating source's type (point/planewave). Research on RIS has been mainly focused on two areas: first, the optimization and design of unit-cells to achieve desired electromagnetic responses within a specific frequency band; and second, exploring the applications of RIS in various settings, including system-level performance analysis. The former does not assume any specific radiation pattern on the surface level, while the latter does not consider any particular unit-cell design. Both approaches largely ignore the complexity and power requirements of the RIS control circuitry. As we progress towards the fabrication and use of RIS in real-world settings, it is becoming increasingly necessary to consider the interplay between the unit-cell design, the required surface-level radiation patterns, the control circuit's complexity, and the power requirements concurrently. In this paper, a benchmarking framework for RIS is employed to compare performance and analyze tradeoffs between the unit-cell's specified radiation patterns and the control circuit's complexity for far-field beamforming, considering different diode-based unit-cell designs for a given surface size. This work lays the foundation for optimizing the design of the unit-cells and surface-level radiation patterns, facilitating the optimization of RIS-assisted wireless communication systems. 
\end{abstract}



\maketitle

\section{Introduction}
Reconfigurable intelligent surface (RIS) allows the reconfiguration of wireless channels~\cite{huang2019reconfigurable,huang2020holographic}. 
The design of an RIS includes a basic planar micro-structure called unit-cell equipped with integrated electronic components such as diodes to allow the tuning of the magnitude and phase of the incident electromagnetic (EM) waves~\cite{di2020smart,zhang2020optically}. A surface is then fabricated by repeating the unit-cells at sub-wavelength periodic intervals. An appropriate control circuit is also added to tune the RIS unit-cells to achieve the desired radiation patterns. An RIS made up of a sufficiently large number of unit-cells can generate complex radiation patterns~\cite{howManyBits2020}. 

\begin{figure*}
    \centering
    \includegraphics[width=1.98\columnwidth]{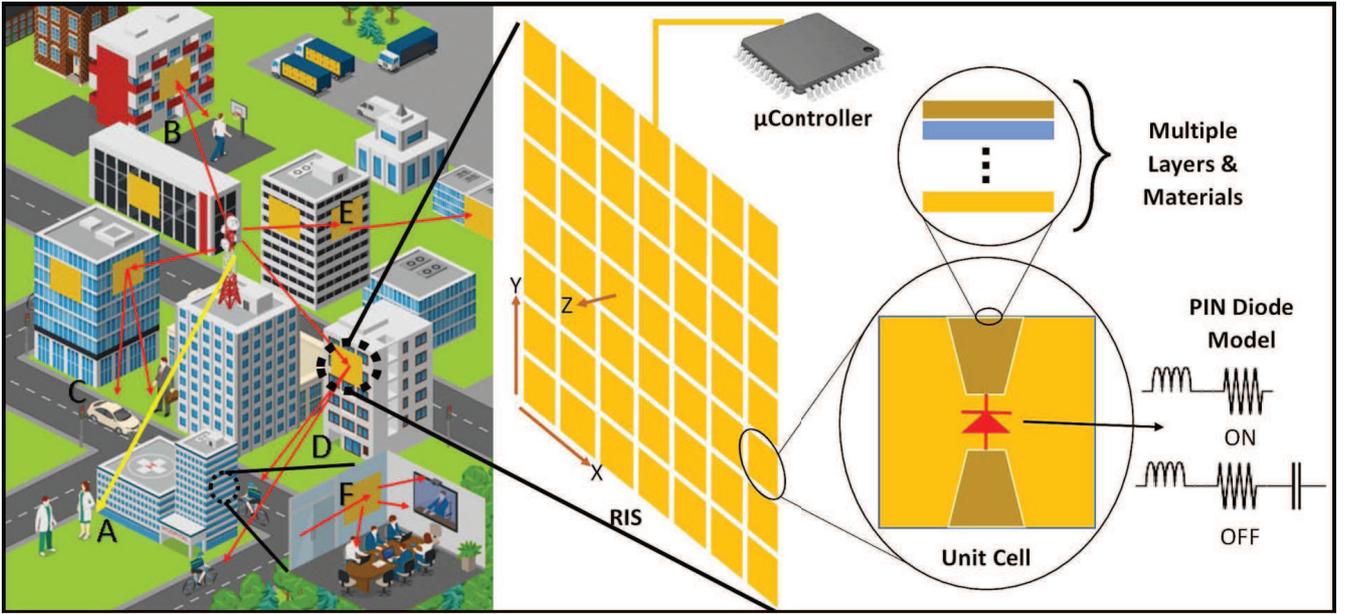} 
    \caption{Example of RIS-assisted urban environment.} 
    \label{fig:smart_env}
\end{figure*}

RIS-assisted wireless communication deployments for an urban environment with indoor/outdoor applications are illustrated in Figure~\ref{fig:smart_env}. The typical structure of a unit-cell, a PIN diode with ON/OFF control, the lumped-element models of the PIN diode, an RIS comprising of multiple unit-cells, and a microcontroller to turn ON/OFF the PIN diodes for the generation of specified radiation patterns are also shown in this figure. Non-line of sight (NLoS) scenarios dominate urban environments where buildings often block the signals. In such situations, single-beam steering, multi-beam forming with equal power levels, and multi-beam forming with unequal power levels can be enabled with the help of RIS to reduce outages and improve the spectral efficiency~\cite{wu2021intelligent_RIS_tutorial,zhang2022beam_distance_angle_resolved}.

The near-field of an antenna or a unit-cell is conventionally defined up to $\frac{2D^2}{\lambda}$ meters far from the antenna or the unit-cell, where $D$ is the minimal diameter of a sphere that encloses the antenna or the unit-cell, and $\lambda$ is the wavelength \cite{balanis2016antenna,bjornson2021primer_whole_building_RIS}. Assuming that the RIS has $N^2$ unit-cells, the near-field region of an RIS is $N^2$ times larger than the near-field boundary of a unit-cell. Depending on the number of unit-cells in an RIS, its near-field region can extend up to tens or hundreds of meters. The users or receivers can, therefore, be located in either the far-field or the near-field region of an RIS. Similarly, an RIS may be located in either the far- or the near-field of a transmitter/source. If the RIS is located in the far-field of a source, the EM waves appear as planewave to RIS, and the source is referred to as a planewave source. On the other hand, if the RIS is located in the near-field of a source, the EM waves appear as spherical waves to the RIS, and the source is referred to as a point source for simplicity. Therefore, in RIS-assisted wireless communications, four cases emerge based on the source type and receiver location from the RIS.
\begin{enumerate}
\item Case 1: The transmitter appears as a \textit{point source} to the RIS and the receiver is located in the \textit{far-field} of the RIS. 
\item Case 2: The transmitter appears as a \textit{planewave source} to the RIS and the receiver is located in the \textit{far-field} of the RIS. 
\item Case 3: The transmitter appears as a \textit{point source} to the RIS and the receiver is located in the \textit{near-field} of the RIS.
\item Case 4: The transmitter appears as a \textit{planewave source} to the RIS and the receiver is located in the \textit{near-field} of the RIS. 
\end{enumerate}

The analytical methods to compute the radiation pattern produced by an RIS differ in each case. The far-field assumption greatly simplifies the computation of the radiation pattern because the elevation and azimuth angles from every unit-cell of the RIS are approximately the same in the far-field~\cite{wei2022codebook_farfield_nearfield_VeryLarge_RIS}. In this paper, we consider the first two cases (case 1 and case 2) and study the ability of an RIS to produce various radiation patterns. This is motivated by the fact that an RIS usually performs better when it is closer to the source \cite{Marco2}, and it may be likely that the receivers are located in the far-field of the RIS.

RIS can play a crucial role in indoor and outdoor scenarios for single-beam steering, multi-beam forming with equal or unequal power levels, and improving the performance of RIS-assisted communication systems, using both point and planewave sources. Various studies have investigated the theoretical gains of RIS in different application settings, as illustrated in Figure \ref{fig:smart_env}, across different frequency ranges. These studies have been comprehensively reviewed in \cite{wu2021intelligent_RIS_tutorial}. However, these theoretical analyses tend to overlook the shape and size of the unit-cells, the complexity and power requirements of the RIS control circuit, and other factors beyond surface-level capabilities. With the increasing use of RIS in practical settings, there is a pressing need to develop tools for selecting appropriate unit-cell designs that align with specific application scenarios using point or planewave sources. The process of optimizing the unit-cells entails considering the trade-offs between unit-cell optimization, surface-level radiation pattern generation capabilities, and control circuit complexity alongside the associated channel sensing overheads. Recently, it was demonstrated in \cite{cheng2021global} that the optimal positioning of RIS in a 3D environment can also enhance the system-level performance of RIS-assisted communication systems. This paper's primary focus is on unit-cell and surface-level RIS design considerations, and it does not delve into system-level optimization.

\begin{figure*}[htb]
 \includegraphics[width=.19\textwidth]{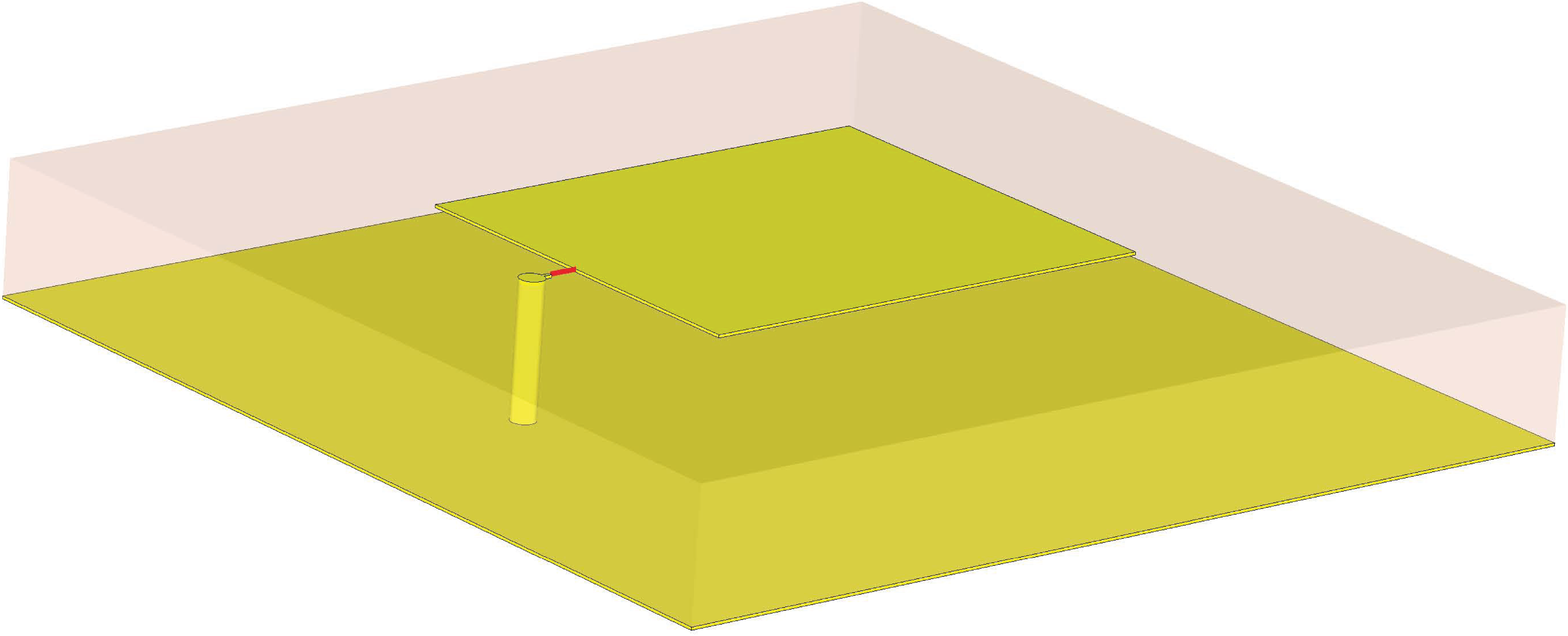}\hfill
 \includegraphics[width=.19\textwidth]{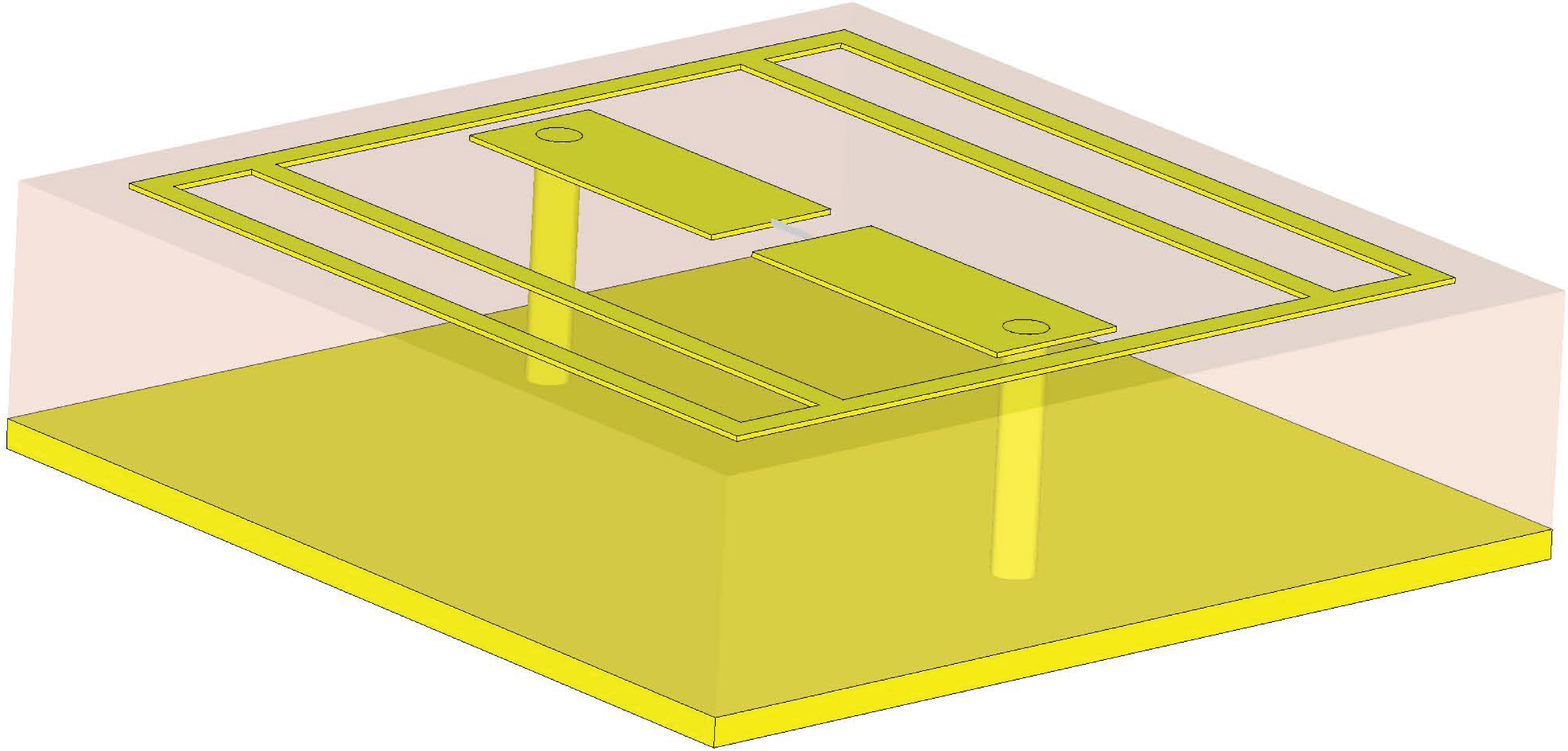}\hfill
 \includegraphics[width=.19\textwidth]{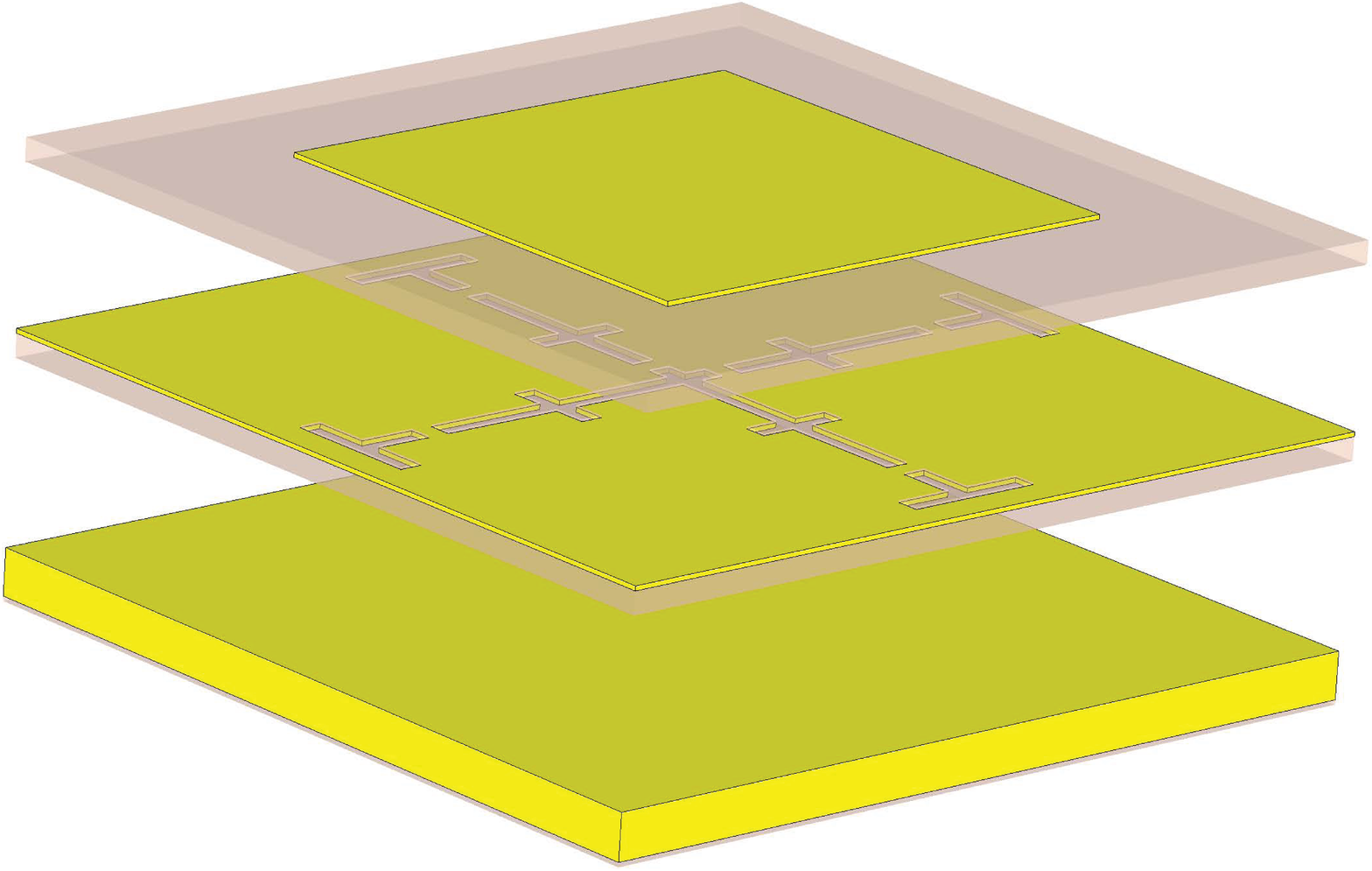}\hfill
 \includegraphics[width=.19\textwidth]{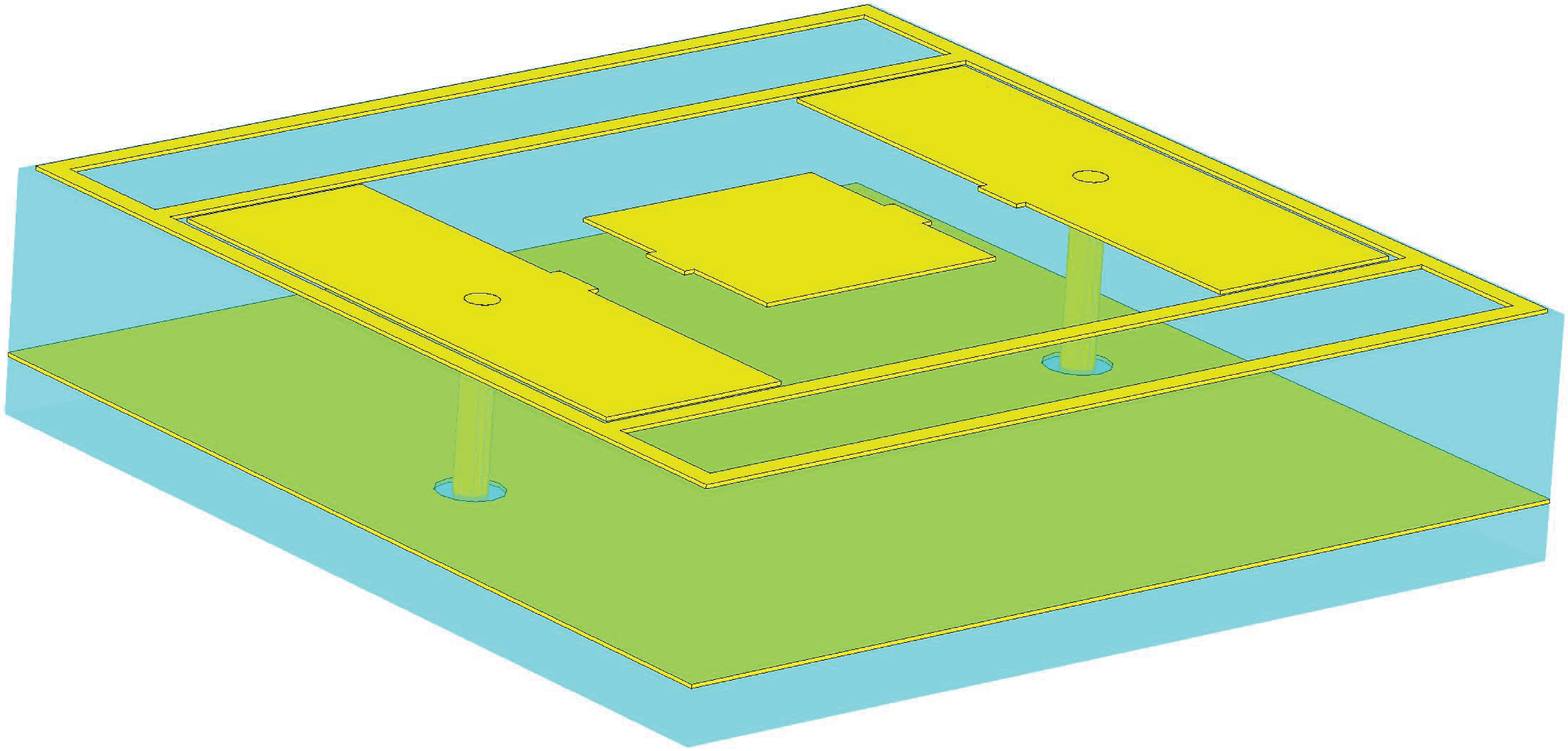}\hfill
 \includegraphics[width=.19\textwidth]{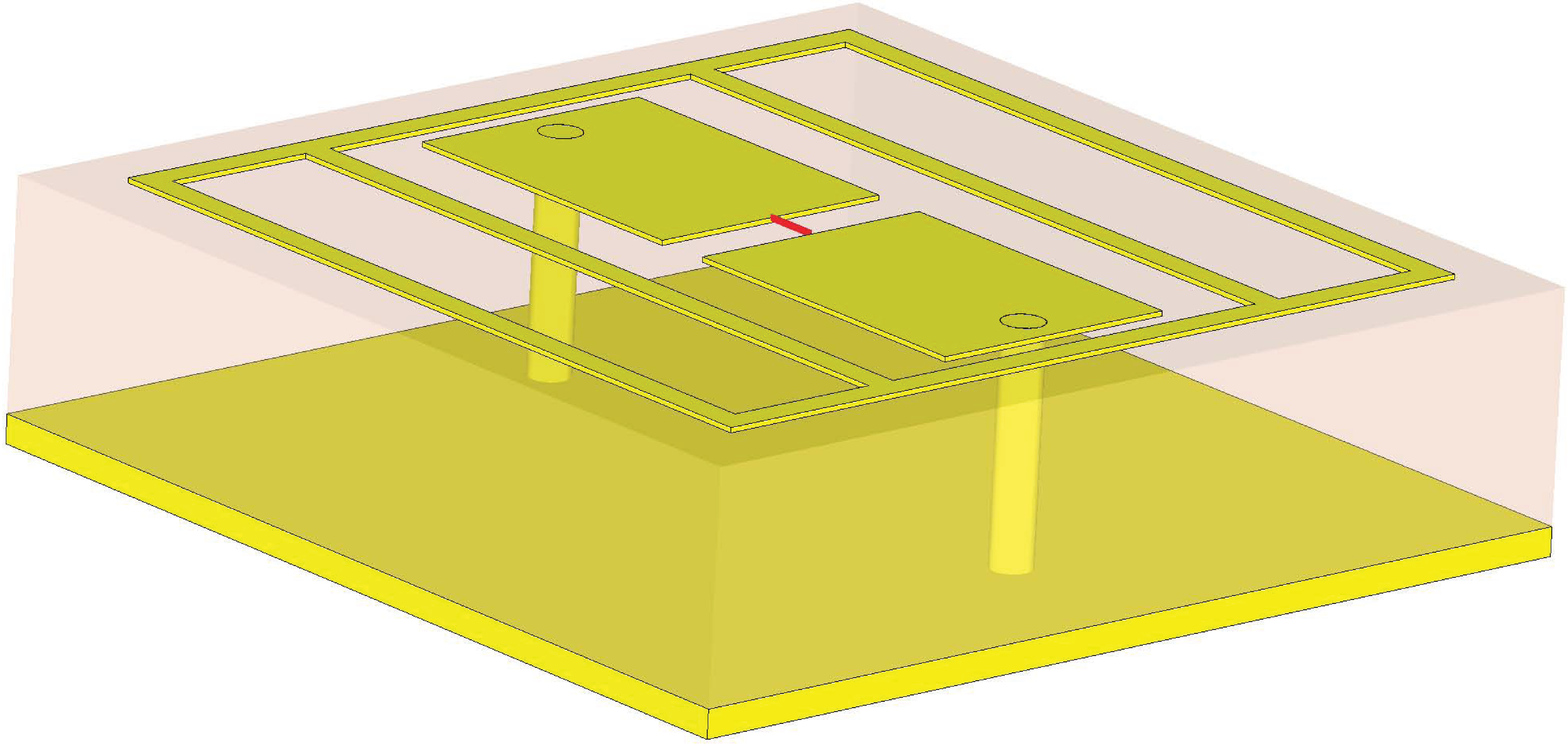}
\caption{({from left to right}) Unit cell layouts of $\text{S}_1$~\cite{yang2016programmable}, $\text{S}_2$~\cite{Xiang2016field}, $\text{S}_3$~\cite{Hanzo2020reconfigurable}, $\text{S}_4$~\cite{Qian2019smart}, $\text{S}_5$~\cite{Xiang2016field}.}
\label{fig:layout}
\end{figure*}

This paper aims to introduce a benchmarking framework and metrics that can be used to assess the radiation pattern generation capabilities, energy requirements, and control circuit complexity of RIS fabricated from various unit-cells, specifically those employing PIN diodes for tuning, under point/planewave source assumptions for far-field beamforming. Additionally, we examine the possibility of trading the radiation pattern generation capability for control circuit complexity through unit-cell grouping. To achieve our goal, we establish a set of radiation patterns, both simple and complex, as benchmarks within our framework because some recent studies emphasize the importance of considering the entire patterns to maximize the power towards the desired directions while keeping it low in the unwanted directions \cite{Marco5}. 

Our study focuses on evaluating the effectiveness of the proposed benchmarking framework and analyzing the tradeoffs for five RISs built using five different unit-cell designs \cite{yang2016programmable, Xiang2016field, Hanzo2020reconfigurable, Qian2019smart}. Among these designs, two RISs utilize 1-bit control achieved through a single PIN diode, while the remaining two RISs utilize 2-bit control via multiple PIN diodes. The unit-cells used in these four RISs are optimized to generate the maximum phase differences between different diode control states, i.e., 180$^{\circ}$ for 1-bit unit-cells and 90$^{\circ}$ for 2-bit unit-cells. The fifth RIS also utilizes a single PIN diode for 1-bit control, but its unit-cell design is unoptimized and results in only a 50$^{\circ}$ phase shift. At normal incidence, all unit-cells in our study exhibit reflection coefficients between 0.85 and 1.

Numerical experiments are conducted to assess the ability of these RISs to replicate the benchmarking patterns in the far-field, assuming a normal angle of incidence for incoming EM waves using point or plane wave sources, which is widely used in the literature \cite{Marco3}. We also analyze and discuss the resulting complexity of the RIS control circuit and power requirements for each RIS. Our key findings are summarized below.
\begin{itemize}
\item When assuming a point source (case 1), we observe that RIS made from a 1-bit unit-cell design outperforms RIS made from a 2-bit design in terms of control circuit complexity and power requirements due to the spherical curvature of the wavefronts.
\item In the case of a planewave source (case 2), RIS made from a 1-bit unit-cell design exhibits significantly poorer beam steering performance compared to RIS made from a 2-bit unit-cell design.
\item When the unit-cells on RIS are controlled in groups, the performance of poor designs is less affected, which also reveals that making large surfaces out of poor designs would hardly improve surface-level performance.
\item If a simple radiation pattern, such as a single beam steering at small reflection angles (less than 40$^{\circ}$), is required, un-optimized unit-cell designs can be used.
\item For generating complex radiation patterns, such as multiple beams at different non-uniform angles, complex and highly-optimized unit-cell designs with multi-bit control and large-sized surfaces are helpful.
\item The power requirements and size of the RIS control circuit increase with the number of PIN diodes per unit-cell design and the operating frequency.
\item RISs are almost passive devices as they do not add new power to the incoming radio signals, but the power requirements of the control circuit cannot be ignored.
\item To effectively compare the generation capabilities of RIS radiation patterns, multiple metrics are required. In this work, we introduce three useful metrics - directivity error (DE), normalized mean squared error (NMSE), and side lobe ratio (SLR) - to quantify the relative performance of various RIS. 

\end{itemize}

The rest of the paper is organized as follows. In Section II, we discuss the unit-cell designs and the electric field (E-field) produced by RIS in the far-field; in Section III, we discuss the RIS control circuit and analyze its complexity and power requirements; in Section IV, we present our benchmarking framework and performance metrics; in Section V, we present our simulation results and their discussion; and in Section VI, we conclude the paper.  

\section{Unit-cell \& RIS}
\label{sec:unit}
\subsection{Unit-Cell \textemdash The basic element of RIS }
A fundamental element of RIS is a planar micro-structure called unit-cell. A flat surface is obtained by arranging unit-cells in rectangular arrays. The unit-cell size depends on the frequency of operation, with higher frequencies requiring smaller dimensions. The total number of unit-cells in a fixed-size RIS depends on its shape and size. By joining multiple smaller surfaces with the repeating pattern of unit-cells, large-sized RIS can be made. It is important to note that we define RIS as a repeating pattern of unit-cells jointly controlled either with a single controller or a set of controllers. 

A category of RIS research focuses entirely on unit-cell design and its EM properties. By including active electronic components, such as PIN diodes, into unit-cell, flexible and real-time functionality is expected from the resulting RIS. Based on these ideas, the RIS made from the unit-cell designed in \cite{yang2016programmable} demonstrates agile scattering, planar focusing, beam steering, and beam forming. The unit-cell proposed in \cite{yang2016programmable} has a sandwich structure composed of a simple rectangular patch, a metal ground plane, and a single PIN diode (1-bit control) connects one edge of the patch to the ground through a metallic via. In \cite{Xiang2016field}, the authors propose a three-layer unit-cell design, which is again controlled through a single PIN diode (1-bit control). A relatively more complex unit-cell is proposed in \cite{Hanzo2020reconfigurable}. This design consists of 5 PIN diodes, but only two control signals are required; therefore, we can classify this as a 2-bit design. Four configurations of 5 PIN diodes produce four almost 90$^{\circ}$ apart phase shifts. The unit-cell is symmetric but has a relatively complex structure and consists of an upper patch, a slot-loaded plane, and a ground. Another 2-bit unit-cell design using only 2 PIN diodes is proposed in \cite{Qian2019smart}. The layouts of all the unit-cells are shown in Figure~\ref{fig:layout}. Table~\ref{tab:sec3table} summarizes the design frequency, reflection amplitudes, and phases of all these unit-cells in different control states at a normal incidence angle. 

In this paper, the RISs made from the unit-cell designs proposed in \cite{yang2016programmable, Xiang2016field, Hanzo2020reconfigurable, Qian2019smart} are denoted as $\text{S}_1$, $\text{S}_2$, $\text{S}_3$, and $\text{S}_4$ respectively. An unoptimized unit-cell design obtained from \cite{Xiang2016field} is also considered and referred to as $\text{S}_5$. This design has a maximum phase shift of only 50$^{\circ}$ between its two configurations, allowing for testing the necessity and extent of unit-cell optimization for achieving good performance. 

\subsection{E-Field of RIS}
In general, the E-field of an RIS comprising of $M\times N$ unit-cells of the same type, arranged in a rectangular planar array, at elevation and azimuth angles denoted by $\theta$ and $\phi$ respectively on the observer side, can be calculated in the far-field as:
\begin{multline}
E(\theta,\phi) = \sum_{m=1}^{M}\sum_{n=1}^{N} \big[E_{in_{mn}}e^{j\alpha_{mn}}f(\theta_{mn}, \phi_{mn})\times\\
\Gamma e^{j\Phi_{mn}} f(\theta,\phi) e^{jk.\hat{r}_{mn}}\big].
\label{eq:array_factor_general}    
\end{multline}
where, $e^{jk.\hat{r}_{mn}}$ represents the wave-vector, $E_{in_{mn}}$ and $\alpha_{mn}$ are the illuminating amplitude and phase (source radiation response), $f(\theta,\phi)$ is the unit-cell radiation response (scattering pattern), the angles $\phi_{mn}$ and $\theta_{mn}$ are the azimuth and elevation angles of the source relative to $(m,n)^{th}$ unit-cell, $\Gamma$ is the reflection coefficient of the unit-cell, and $\Phi_{mn}$ is the phase shift produced by $(m,n)^{th}$ unit-cell. This phase shift is controlled by changing the state of the PIN diodes. For example, for an optimized 2-bit unit-cell the value of $\Phi_{mn} \in \{ 0, \pi/2, \pi, 3\pi/2\}$. Thus, by varying the configuration of unit-cells, different responses can be achieved by RIS, i.e., it can steer the incoming signal in the arbitrary desired direction(s). 

Equation \eqref{eq:array_factor_general} is valid for both case 1 and case 2. However, under the planewave source assumption (case 2), we can further simplify some terms in this equation. The radiation response from the source simplifies to $E_{in_{mn}}=E, \: \forall m,n$, and $e^{j\alpha_{mn}}=1, \: \forall m,n$ (or some constant). Similarly, the unit-cell radiation response becomes $f(\theta_{mn}, \phi_{mn})=f(\theta,\phi)$ because all the azimuth and elevation angles of the source are nearly equal, i.e., $\theta_{mn}\approx \theta, \: \forall m,n$ and $\phi_{mn}=\phi, \: \forall m,n$. Suppose we further assume that RIS lies in the x-y plane and unit-cell radiates equally on all sides in the plane. In that case, the unit-cell radiation response further simplifies to $f(\theta), \: \forall m,n$ (there is no dependence on angle $\phi$). With all these simplifications, the E-field equation for case 2 becomes:
\begin{equation}
E(\theta,\phi) = E \: \Gamma \: f^2(\theta) \sum_{m=1}^{M}\sum_{n=1}^{N} e^{j\Phi_{mn}} e^{jk.\hat{r}_{mn}}.
\label{eq:case2}    
\end{equation}

When we compare \eqref{eq:array_factor_general} and \eqref{eq:case2}, it becomes evident that the planewave source assumptions limit the options for beam manipulation to a relatively lesser degree, however, this simplifies the E-field computations because the unit-cell radiation response $f(\theta)$ remains unchanged for all indices $(m,n)$, and it can be factored out as a constant from the summations.

\begin{figure}
    \centering
    \includegraphics[width=\columnwidth]{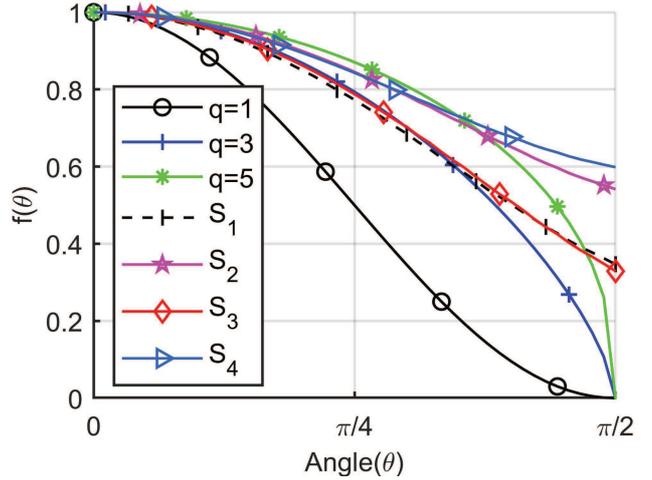} 
    \caption{Normalized radiation response $f(\theta)$ of the four unit-cells used in $\text{S}_1$-$\text{S}_4$. The powers of $\cos^{\frac{1}{q}}(\theta)$ for $q=1,3,5$ are also plotted.} 
    \label{fig:normalized_radiation_patterns}
\end{figure}

The design of the unit-cell generally influences the shape of $f(\theta)$. Figure \ref{fig:normalized_radiation_patterns} displays the normalized radiation response of the four unit-cells studied in this paper and also plots $f(\theta)=\cos^{\frac{1}{q}}(\theta)$ for certain values of $q$ for comparison. The radiation response of the unit-cells used in $\text{S}_1$ and $\text{S}_3$ is similar to $\cos^{\frac{1}{3}}(\theta)$, while that of $\text{S}_2$ and $\text{S}_4$ is similar to $\cos^{\frac{1}{5}}(\theta)$. However, assuming a planewave source, a sufficiently large-sized RIS constructed using any $n$-bit optimized unit-cell with $\Gamma=1$ and maximally separated phase shifts between diode states should produce almost similar surface level performance, regardless of the differences in $f(\theta)$. In other words, the number of unit-cell diode states ($2^n$) plays a more critical role in determining the surface-level performance of the RIS than the unit-cell radiation response $f(\theta)$ does. As $n$ increases, the RIS's ability to generate relatively complex radiation patterns grows, as does the control circuit's complexity and power requirements (discussed in more detail in the following section). However, under point source assumptions, the shape of $f(\theta)$ plays a more significant role in determining the surface level response, as it appears inside the summation in \eqref{eq:array_factor_general}. Thus, in case 1, we should anticipate considerable surface-level performance variations based on the unit-cell design.

There is a lack of standardized benchmarks to evaluate and compare the radiation pattern generation abilities of RIS made of different unit-cells, their control circuit complexity, and power requirements. Moreover, various authors assume different numbers of RIS elements and spacing between them, with some using 40x40, 20x20, 16x16, or 30x30 unit-cells in their studies. Such differences make it challenging to compare the relative performance of different unit-cell designs in generating surface-level radiation patterns under point/planewave source assumptions. 
In practical applications, the conventional design approach involves selecting or designing a unit-cell structure, including the control mechanism (PIN diodes), and then using multiple unit-cells to create a surface capable of generating specific radiation patterns. However, this method has limitations, as the resulting finite-sized RIS may not possess the necessary pattern generation capabilities, may not be suitable for the intended application, or may consume excessive power.

A significant difficulty also lies in finding the states of the PIN diodes of each unit-cell to achieve some desired radiation response. The search space has exponential complexity, and the problem is NP hard\cite{yang2016programmable} even for a moderately sized RIS comprising 20 to 30 1-bit unit-cells. While efficient algorithms exist to determine the states of the PIN diodes, they primarily target wireless communications and rely on simplistic models. Data rates at specific locations are their primary design objective rather than the entire radiation pattern. Thus, these algorithms do not account for how an RIS reflects signals in other directions during the design stage. It is crucial to acknowledge that even the most efficient algorithms may not succeed if a finite-sized RIS created from a given unit-cell design cannot generate a specific radiation pattern that was not previously tested. These difficulties motivate the need for helpful radiation patterns that can act as benchmarks to test and compare the performance of various designs.  

\section{Control and Power Requirements of RIS}
\label{sec:control}
This section discusses the control circuit complexity and power requirements of RIS built from different unit-cell designs. Table~\ref{tab:sec3table} summarizes the control circuit complexity, power requirements, and unit area power requirements of RIS $\text{S}_1$-$\text{S}_5$.  

\begin{figure}
    \centering
    \includegraphics[width=\columnwidth]{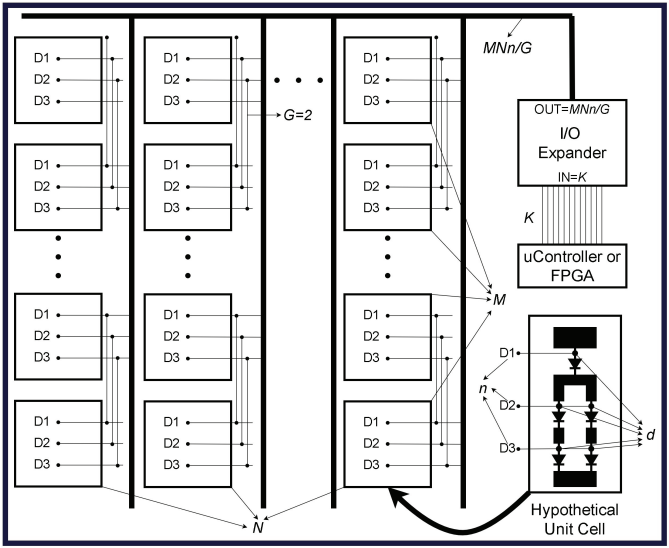}
    \caption{RIS control circuit, unit-cell grouping and I/O expansion for addressing.}
    \label{fig:RIScontrol}
\end{figure}

\subsection{Physical and Logical Control Paths on RIS}
Each unit-cell incorporates PIN diodes and therefore requires independent control signals. For an $n$-bit unit-cell, $n$ separate control lines are required, and $2^n$ distinct phases can be generated. It should be noted that the total number of PIN diodes in an $n$-bit unit-cell may exceed $n$, as in the case of~\cite{Hanzo2020reconfigurable}, where five PIN diodes are used to produce four distinct phase shifts with only two control signals. The control signals are generated by a microcontroller or FPGA, which requires an additional interface circuit. However, these controllers have a limited number of pins, which may not be sufficient to control all the unit-cells on a large RIS simultaneously. To address this issue, time division multiplexing is typically used, and this control architecture type is referred to as active matrix architecture or I/O expansion~\cite{kamoda201160}.

Compared to varactor diodes, control circuits made of PIN diodes are more straightforward, and digital high and low
states are sufficient for their ON/OFF control~\cite{lim2010reconfigurable_pind_varactor}. Additionally, PIN diodes have a much lower forward biasing voltage (0.7V--0.8V), but the forward-biased current is of the order of mA. Conversely, varactors operate at a significantly higher voltage than a digital high state of a PIN diode but require $\mu$A or less current when fabricated with the same technology~\cite{Marco6}. As a result, varactors generally consume less power than PIN diodes. Nonetheless, PIN diodes are preferred because of their control simplicity.

The rate at which an RIS can switch between different functionalities (we refer to it as the RIS function switching)
is directly dependent on the speed of the control circuit. In fast-changing wireless channels, the RIS function switching must also be fast. However, the number of unit-cells on the RIS affects the channel sensing overhead and the complexity of the control circuit. Fewer unit-cells mean a lower channel sensing overhead and less complex control circuit, but it could affect the RIS's ability to generate complex radiation atterns~\cite{howManyBits2020}. To reduce channel sensing overhead and control circuit complexity, unit-cells can be grouped and controlled simultaneously. This also negatively impacts the RIS's radiation pattern generation capabilities and overall functionality. To investigate the extent to which unit-cell grouping affects the performance of RIS S1-S5, it is necessary to quantify these tradeoffs using the benchmarking framework and performance metrics discussed in the next section.

Using Figure~\ref{fig:RIScontrol}, we illustrate the concept of unit-cell grouping and distinguish between physical and logical control paths. The figure depicts groups of two unit-cells, where each unit-cell contains three PIN diodes labeled as D1, D2, and D3, enclosed within a rectangular box. The RIS comprises $MN$ unit-cells, with $K$ microcontroller pins available for control purposes, where $K < MN$. The $K$ pins drive control signals to the first $K$ unit-cells or groups of unit-cells at the same time, while the remaining cells or groups are disabled. Then, the next $K$ unit-cells or groups of unit-cells are enabled, and the same $K$ pins provide new control signals. This sequence continues until all the unit-cells or groups of unit-cells have been addressed, leading to logically separated control paths. It is worth noting that increasing the number of controllers that operate in parallel can increase the overall design cost but also allow for an increase in the total number of control pins.

The number of physical control circuit paths can be expressed as $\frac{MNn}{G}$, where $MN$ is the total number of unit-cells, and $G$ is the number of unit-cells in each group. By increasing the size of the unit-cell groups, we can significantly reduce the complexity of the control circuits. The order of the RIS function switching rate can be quantified as $O \left(\frac{GK}{MNn \tau} \right)$, where $\tau$ represents the response time of the slowest element in the control path. For example, if we consider an RIS composed of 2-bit unit-cells with $G=2$, $M=N=40$, $K=40$, and $\tau=20$ns, the function switching rate would be 1.25 MHz, enabling the RIS to switch from one radiation pattern to another in approximately 0.8$\mu$s. In the frequency range of 1 GHz to 30 GHz, the typical wireless channel coherence time varies from a few hundred $\mu$s to a few $\mu$s, which implies that the RIS can quickly adapt to changing wireless channel conditions. Additionally, several PIN diodes, including SMP1340, can operate at a GHz switching rate, which can make the RIS function switching even faster. However, the actual switching rate would depend on several factors, such as the channel sensing overhead and the power requirements of the switching circuitry. We can also observe a tradeoff between the number of physically independent control circuit paths and the function switching rate.

\begin{table*}[htb]
\centering
\caption{Control circuit complexity and maximum power requirements of RIS $\text{S}_1$-$\text{S}_5$ each made from $MN$ unit-cells. $P_D=8mW$.}
\renewcommand{\arraystretch}{1.3}
\begin{tabular}{|l|l|l|cc|lclllc|}
\hline
\multicolumn{1}{|c|}{\multirow{2}{*}{\textbf{RIS}}} &
  \multicolumn{1}{c|}{\multirow{2}{*}{\textbf{$n$}}} &
  \multicolumn{1}{c|}{\multirow{2}{*}{\textbf{$d$}}} &
  \multicolumn{2}{c|}{\textbf{\begin{tabular}[c]{@{}c@{}}RIS Control Circuit \\ Complexity \end{tabular}}} &
  \multicolumn{6}{c|}{\textbf{Maximum RIS Power Requirements}} \\ \cline{4-11} 
\multicolumn{1}{|c|}{} &
  \multicolumn{1}{c|}{} &
  \multicolumn{1}{c|}{} &
  \multicolumn{1}{c|}{\textbf{\begin{tabular}[c]{@{}c@{}}RIS Physical \\ Control \\ Circuit \\ Paths\end{tabular}}} &
  \textbf{\begin{tabular}[c]{@{}c@{}}RIS Function \\ Switching \\ Rate\end{tabular}} &
  \multicolumn{1}{c|}{\textbf{\begin{tabular}[c]{@{}c@{}}RIS Total \\ Power \\ $(dMNP_D)$\end{tabular}}} &
  \multicolumn{1}{c|}{\textbf{\begin{tabular}[c]{@{}c@{}}Unit-Cell \\ Dimensions \\ (mm)\end{tabular}}} &
  \multicolumn{1}{c|}{\textbf{\begin{tabular}[c]{@{}c@{}}Unit \\ Cell \\ Area  \\ $\times10^{-4}$\\(m$^2$)\end{tabular}}} &
  \multicolumn{1}{c|}{\textbf{\begin{tabular}[c]{@{}c@{}} Reflection \\ Coefficient \\ at normal incidence angle \\ (magnitude$\angle$phase$^\circ$)   \end{tabular}}} &
  \multicolumn{1}{c|}{\textbf{\begin{tabular}[c]{@{}c@{}} $f$ (GHz) \end{tabular}}} &
  \multicolumn{1}{c|}{\textbf{\begin{tabular}[c]{@{}c@{}}RIS \\ Power \\Per \\ Unit \\ Area \\ (W/m$^2$)\end{tabular}}} \\ \hline
$\text{S}_1$\cite{yang2016programmable} &
  1 &
  1 &
  \multicolumn{1}{c|}{$MN/G$} &
  $GK/(MN\tau)$ &
  \multicolumn{1}{c|}{$MNP_{D}$} &
  \multicolumn{1}{c|}{$5.8 \times 4.9 $} &
  \multicolumn{1}{c|}{$1.82$} &
  \multicolumn{1}{c|}{\begin{tabular}[c]{@{}c@{}} 0.95$\angle_1 0^\circ$, 0.92$\angle_2 180^\circ$ \end{tabular}} &
  \multicolumn{1}{c|}{11.1} &
  44 \\ \hline
$\text{S}_2$\cite{Xiang2016field} &
  1 &
  1 &
  \multicolumn{1}{c|}{$MN/G$} &
  $GK/(MN\tau)$ &
  \multicolumn{1}{c|}{$MNP_{D}$} &
  \multicolumn{1}{c|}{$6 \times 6$} &
  \multicolumn{1}{c|}{$2.89$} &
  \multicolumn{1}{c|}{\begin{tabular}[c]{@{}c@{}} 0.90$\angle_1 0^\circ$, 0.88$\angle_2 180^\circ$ \end{tabular}} &
  \multicolumn{1}{c|}{8.82} &
  27.7 \\ \hline
$\text{S}_3$\cite{Hanzo2020reconfigurable} &
  2 &
  5 &
  \multicolumn{1}{c|}{$2MN/G$} &
  $GK/(2MN\tau)$ &
  \multicolumn{1}{c|}{$5MNP_{D}$} &
  \multicolumn{1}{c|}{$50 \times 50$} &
  \multicolumn{1}{c|}{$42.3$} &
  \multicolumn{1}{c|}{\begin{tabular}[c]{@{}c@{}} 0.95$\angle_1 0^\circ$, 0.95$\angle_2 90^\circ$, \\ 0.98$\angle_3 180^\circ$, 0.92$\angle_4 270^\circ$ \end{tabular}} &
  \multicolumn{1}{c|}{2.3} &
  9.46 \\ \hline
$\text{S}_4$\cite{Qian2019smart} &
  2 &
  2 &
  \multicolumn{1}{c|}{$2MN/G$} &
  $GK/(2MN\tau)$ &
  \multicolumn{1}{c|}{$2MNP_{D}$} &
  \multicolumn{1}{c|}{$8.8 \times 8.8$} &
  \multicolumn{1}{c|}{$2.72$} &
  \multicolumn{1}{c|}{\begin{tabular}[c]{@{}c@{}} 0.88$\angle_1 0^\circ$, 0.85$\angle_2 90^\circ$, \\ 0.92$\angle_3 180^\circ$, 0.90$\angle_4 270^\circ$ \end{tabular}} &
  \multicolumn{1}{c|}{9.08} &
  58.8 \\ \hline
$\text{S}_5$\cite{Xiang2016field} &
  1 &
  1 &
  \multicolumn{1}{c|}{$MN/G$} &
  $GK/(MN\tau)$ &
  \multicolumn{1}{c|}{$MNP_{D}$} &
  \multicolumn{1}{c|}{$6 \times 6$} &
  \multicolumn{1}{c|}{$6.25$} &
  \multicolumn{1}{c|}{\begin{tabular}[c]{@{}c@{}} 0.92$\angle_1 0^\circ$, 0.94$\angle_2 50^\circ$ \end{tabular}} &
  \multicolumn{1}{c|}{6.00} &
  12.8 \\ \hline
\end{tabular}
\label{tab:sec3table}
\end{table*}

\subsection{Power Requirements of RIS Control Circuit}
The power consumption of a PIN diode when in the ON state depends on its forward voltage drop and forward operating current. While the power consumption of a single PIN diode is relatively low, typically around 7-8mW~\cite{lim2010reconfigurable_pind_varactor}, it becomes significant when multiple unit-cells with multiple PIN diodes are combined to form large surfaces. It is challenging to determine the average percentage of unit-cells that need to be forward-biased to achieve various functionalities. However, we know that the power requirements of an RIS are directly proportional to the number of PIN diodes on its surface. The proportionality constant is the ratio of PIN diodes in the ON state to the total number of PIN diodes. To simplify and facilitate comparisons, we assume that this proportionality constant is equal to 1\footnote{In reality, the proportionality constant should be less than 1 because only a fraction of diodes would be in the ON state for producing typical radiation patterns.}.

The maximum power requirement of an RIS made consisting of $MN$ unit-cells is $dMNP_D$, where $P_D$ is the power used by one PIN diode when forward-biased (in the ON state). It is apparent that a unit-cell configuration where $d>n$ is less efficient than one where $d=n$. Additionally, as the value of $n$ per unit-cell increases, so does the power requirement. The impact of grouping the unit-cells on power usage would rely on the diode states for each group. For instance, if we take the 1-bit unit-cell design, all of the unit-cells in a group would be ON if they received an ON signal. Conversely, if they receive an OFF signal, all of them would be OFF. The maximum power requirements, on average, would remain the same. However, grouping simplifies the control circuit since all of the unit-cells in a group will be in the same state, based on a single control signal.

Supplying power continuously to an RIS that consists of several hundred unit-cells operating at very high frequencies can pose a challenge. The power consumption of RIS per unit area (W/m$^2$) is a crucial factor to consider. In Table~\ref{tab:sec3table} we have computed these requirements for RIS $\text{S}_1$-$\text{S}_5$, assuming that the unit-cells are arranged uniformly half wavelength apart. The Table also presents the unit-cell dimensions and design frequencies. For example, RIS-$\text{S}_1$ operates at a frequency of 11.1~GHz, and its power consumption per unit area is 44W/m$^2$. Suppose we want to use solar panels that produce between 100-200W/m$^2$ during the sunshine hours. With this assumption, the size of solar panels required to power up RIS-$\text{S}_1$ is less than the size of the RIS. However, these observations quickly change when we consider higher frequencies (above 30 GHz), because the unit-cell dimensions become small, and the size of solar panels required to power up the resulting RIS becomes greater than the RIS size itself. 

\section{Benchmarking Framework - RIS Radiation Patterns and Performance Metrics}
\label{sec:bench}
In a smart city environment where NLOS situations dominate, we expect RIS to possess three significant functionalities. These are single beam steering, multi-beam forming with equal power levels, and multi-beam forming with different power levels. In the following subsection, we propose a flexible benchmarking framework that allows us to test these beamforming capabilities for RIS made from different unit-cell designs under point/planewave source assumptions (case 1 and case 2). 

\subsection{Benchmark Radiation Patterns}
\begin{figure*}
    \centering
    \includegraphics[width=0.98\textwidth]{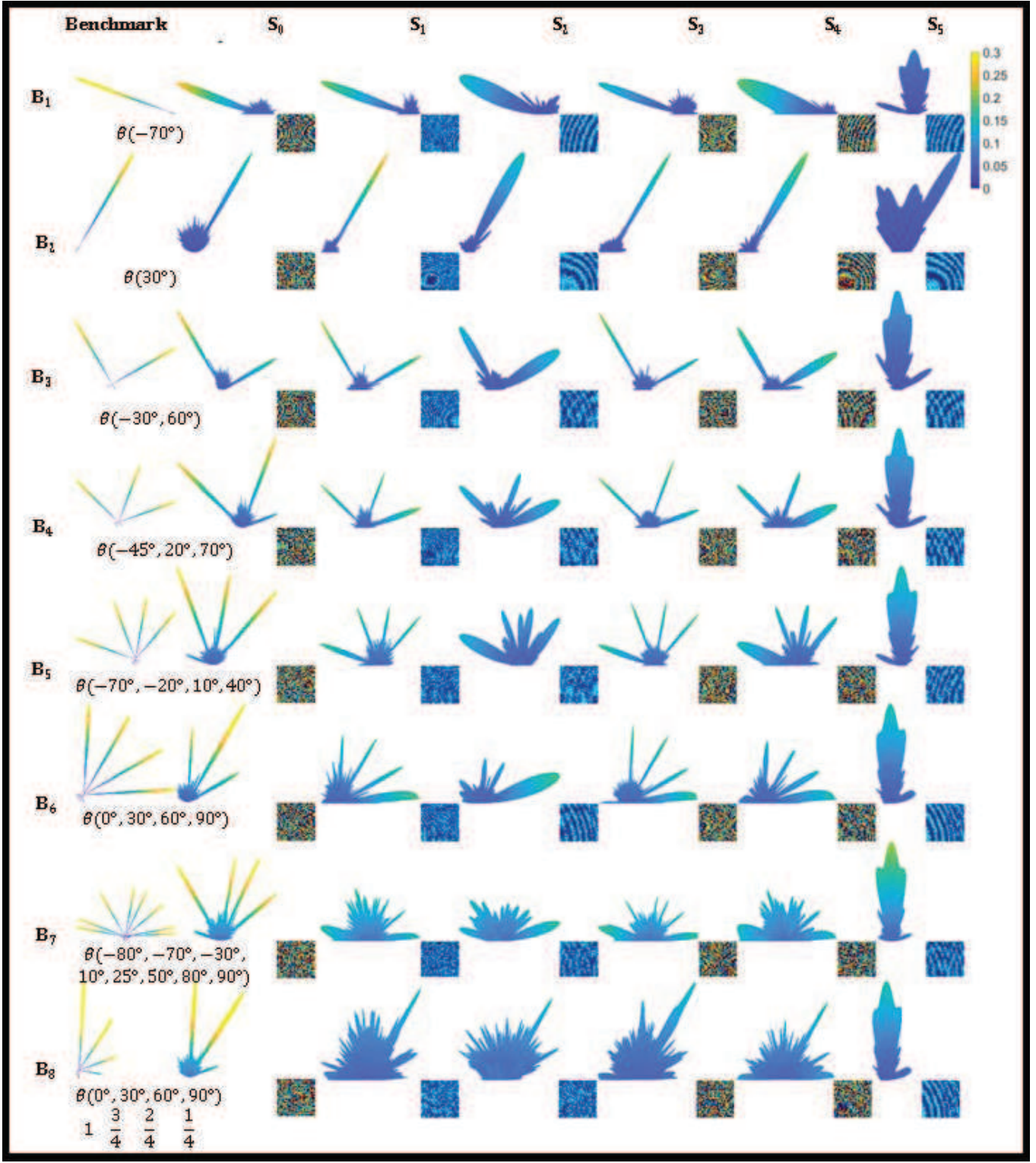}
    \caption{Benchmarking radiation patterns $\text{B}_1$-$\text{B}_8$ and the radiation patterns obtained by RIS $\text{S}_1$-$\text{S}_5$ under point source assumptions (case 1) without unit-cell grouping. The results for the reference RIS S$_0$ are also shown. The RIS configuration matrix ($40\times40$) is also reported as a colored image (unit-cell state 1=Blue, 2=Cyan, 3=Yellow, 4=Red). The elevation identifies Beam directions ($\theta$, ranges between 0$^\circ$ and 90$^\circ$) and azimuth($\phi$, only takes two values 0$^\circ$ and 180$^\circ$) angles. Instead of repeating the two $\phi$ values, $\theta$ and $-\theta$ are used ($-\theta = \theta_{\phi=180^\circ}$).}
    \label{fig:benchmark_plots}
\end{figure*}

As benchmarks, we determine a set of typical radiation patterns based on NLOS scenarios. The considered set includes eight radiation patterns, which are shown in the first column of Figure~\ref{fig:benchmark_plots}. These radiation patterns are labeled as $\text{B}_1$-$\text{B}_8$ for referencing. Radiation patterns $\text{B}_1$ and $\text{B}_2$ are used to test the beam steering capabilities of RIS in a single direction. Radiation patterns $\text{B}_3$ (2 beams), $\text{B}_4$ (3 beams), $\text{B}_5$ (4 beams), $\text{B}_6$ (4 beams), and $\text{B}_7$ (8 beams) are used to test multi beamforming capabilities with equal power. Radiation pattern $\text{B}_8$ is used to test the formation of multiple unequal power beams (4 beams). The relative angles of various beams in these radiation patterns are shown in Figure~\ref{fig:benchmark_plots}. These patterns are referred to as the benchmarking patterns in further discussion. The prominent beams in these benchmarking patterns will be referred to as the main lobes. The side lobes in all the benchmarking patterns are desired to be negligible. In our study, the polarization of the incident EM wave and the operating frequency are assumed to match the design assumptions of each unit-cell in the original papers. Our proposed framework is flexible, and more radiation patterns can be included in the benchmarking set depending on the application scenarios. 

\subsection{Performance Metrics}
We are interested in quantifying the relative ability of different RISs in reproducing the benchmarking patterns. To this end, appropriate metrics that can compare two radiation patterns are required. In this discussion, we will use the terms `achieved radiation pattern' and `reference radiation pattern,' where achieved radiation pattern means the pattern produced by RIS in response to a given reference radiation pattern. We develop three performance metrics. The first metric is the directivity error (DE) which measures the ability of RIS to reproduce the main lobes of the reference radiation pattern. The second metric is the normalized mean squared error (NMSE) which quantifies the accumulative mean square error between the achieved and reference radiation patterns in all directions. The third metric is the side lobe ratio (SLR) which quantifies the main lobe to side lobe ratio where the location of the main lobe is determined from the reference radiation pattern. Please note that our DE, NMSE, and SLR are comparative metrics, i.e., we use them to check how closely the achieved radiation pattern matches a reference radiation pattern. Below we further explain these metrics. \\

\noindent \textbf{DE}: We define DE in the following way: 
\begin{equation}
\text{DE} = \frac{D_r - D_a}{D_r}.
    \label{eq:dirErrCorrect}
\end{equation}
where, $D_r$ is the reference directivity value that is determined by integrating power (square of E-field) over the beam-width of the main lobe in the reference radiation pattern, while $D_a$ is the achieved directivity value that is determined by integrating power over the starting and ending angle values obtained from the beam-width of the main lobe of the reference radiation pattern in the achieved radiation pattern. We calculate the beam-width of the main lobe using the First null beam-width (FNBW) criteria. Please note that in both $D_r$ and $D_a$, starting and ending angles of the main lobe of the reference radiation pattern are used because the objective is to reproduce the main lobe of the reference pattern. The value of DE should generally be positive with a maximum of 1 due to normalization by $D_r$. The DE value of 1 would indicate that the main lobes are not formed in the intended directions and are in completely different directions (inferior performance). The negative value of DE would mean that the achieved radiation pattern is better than the target radiation pattern (this is possible if benchmarking pattern is not used as a reference radiation pattern).  \\
\\ \noindent \textbf{NMSE}: The NMSE is computed as:
\begin{equation}
\text{NMSE} = \frac{1}{L}\sum_{\theta,\phi}
\bigg(\frac{E_r(\theta,\phi)}{E_{rmax}} - \frac{E_a(\theta,\phi)}{E_{amax}} \bigg)^2
\label{eq:nmseCorrect}
\end{equation}
where, $E_r(\theta,\phi)$ is the E-field value of reference radiation pattern and $E_a(\theta,\phi)$ is the E-field value of the achieved radiation pattern at azimuth and elevation angles $\theta$ and $\phi$. We normalize the E-field values of each radiation pattern by the corresponding maximum values, i.e., $E_{rmax}$ for the reference radiation pattern values and $E_{amax}$ for the achieved radiation pattern values. We then average the squared difference of the normalized E-field values in all directions to obtain NMSE. In the simulations, we use $L=180 \times 360$. \\
\\ \noindent \textbf{SLR}: SLR is computed in dB scale:
\begin{equation}
\text{SLR} = 10\log_{10}\frac{\text{Power density of intended lobe}}{\text{Power density of side lobe}}
    \label{eq:sllCorrect}
\end{equation}
Please note that we determine the intended lobe's location (starting and ending angles) from the reference radiation pattern. At these angles, we look for the power density in the achieved radiation pattern. On the other hand, we determine the side lobe exclusively from the achieved radiation pattern of RIS and define it as the most prominent non-intended minor lobe. We are using the words `intended' because these metrics aim to measure how faithfully a reference radiation pattern is reproduced.  
In our set, we also have some multi-beam (B$_{3}-\text{B}_8$) reference radiation patterns. For such patterns, we individually compute the SLR for each intended beam according to \eqref{eq:sllCorrect} and then report their average.  

The benchmarking patterns in our set are ideal, with negligible non-intended lobes. Any RIS of suitable size made from 1-bit or 2-bit unit-cells would struggle to reproduce these patterns faithfully. Therefore, considering the benchmarking patterns as reference radiation patterns for the computation of the quantitative metrics is not an effective way to analyze the relative performance of S$_1$-S$_5$. To avoid this problem, we can select some appropriate unit-cell and then make a reference RIS. We refer to such a reference RIS as $\text{S}_0$. We can compute the quantitative metrics with respect to the radiation patterns produced by $\text{S}_0$.
Please note that for computing the performance metrics, the starting and ending angles of the main lobes will always be those given in the benchmarking patterns, even when we are using the reference RIS method. With this approach, the values of DE could also become negative, and it would indicate that the performance of given RIS in producing the beams in the intended directions, as shown in the benchmarking pattern, is better than that of S$_0$.  

For performance quantification in the next section, we will consider a reference 2-bit unit-cell whose normalized power radiation pattern $f(\theta)$ will be $\cos(\theta)$ \cite{Marco4,Marco1}. This response is plotted in Figure~\ref{fig:normalized_radiation_patterns} as $q=1$ curve. We will assume that this unit-cell is perfectly optimized with $\Gamma=1$ and the phase differences between any two consecutive diode states are 90$^{\circ}$. The RIS made from this reference unit-cell will be called $\text{S}_0$, and we will use it to evaluate the performance metrics. 

\begin{table*}[t]
\centering
\caption{Quantitative results of five RIS S$_{1}-\text{S}_5$, with and without unit-cell grouping under point source assumptions with the observer in the far-field (case 1). Best values (smallest in case of DE and NMSE and largest in case of SLR) are highlighted in \textbf{bold} font, worst values (largest in case of DE and NMSE and smallest in case of SLR) are highlighted in \textit{italic} font.}
\resizebox{\textwidth}{!}{
\renewcommand{\arraystretch}{1.3}
\begin{tabular}{|c|c|lllll||lllll||lllll|}
\hline
ID &
  Grouping &
  \multicolumn{5}{c|}{Relative Directivity Error (DE)} &
  \multicolumn{5}{c|}{Relative Normalized Mean Squared Error (NMSE)} &
  \multicolumn{5}{c|}{Side Lobe Ratio (SLR) - dB} \\ \hline
 &
   &
  \multicolumn{1}{c|}{$S_1$} &
  \multicolumn{1}{c|}{$S_2$} &
  \multicolumn{1}{c|}{$S_3$} &
  \multicolumn{1}{c|}{$S_4$} &
  \multicolumn{1}{c|}{$S_5$} &
  \multicolumn{1}{c|}{$S_1$} &
  \multicolumn{1}{c|}{$S_2$} &
  \multicolumn{1}{c|}{$S_3$} &
  \multicolumn{1}{c|}{$S_4$} &
  \multicolumn{1}{c|}{$S_5$} &
  \multicolumn{1}{c|}{$S_1$} &
  \multicolumn{1}{c|}{$S_2$} &
  \multicolumn{1}{c|}{$S_3$} &
  \multicolumn{1}{c|}{$S_4$} &
  \multicolumn{1}{c|}{$S_5$} \\ \hline
$\text{B}_1$ &
  \multirow{8}{*}{No} &
  \multicolumn{1}{l|}{0.5} &
  \multicolumn{1}{l|}{0.39} &
  \multicolumn{1}{l|}{\textbf{0.16}} &
  \multicolumn{1}{l|}{\textit{1}} &
  0.94 &
  \multicolumn{1}{l|}{0.1004} &
  \multicolumn{1}{l|}{\textit{0.1248}} &
  \multicolumn{1}{l|}{0.091} &
  \multicolumn{1}{l|}{0.1131} &
  \textbf{0.0889} &
  \multicolumn{1}{l|}{11.58} &
  \multicolumn{1}{l|}{9.02} &
  \multicolumn{1}{l|}{8.69} &
  \multicolumn{1}{l|}{\textbf{12.62}} &
  \textit{-1.11} \\
$\text{B}_2$ &
   &
  \multicolumn{1}{l|}{0.3} &
  \multicolumn{1}{l|}{0.54} &
  \multicolumn{1}{l|}{0.7} &
  \multicolumn{1}{l|}{\textbf{0.24}} &
  \textit{0.82} &
  \multicolumn{1}{l|}{\textbf{0.0284}} &
  \multicolumn{1}{l|}{0.0459} &
  \multicolumn{1}{l|}{0.0381} &
  \multicolumn{1}{l|}{0.0311} &
  \textit{0.0516} &
  \multicolumn{1}{l|}{9.96} &
  \multicolumn{1}{l|}{11.01} &
  \multicolumn{1}{l|}{8.86} &
  \multicolumn{1}{l|}{\textbf{11.62}} &
  \textit{6.27} \\
$\text{B}_3$ &
   &
  \multicolumn{1}{l|}{0.22} &
  \multicolumn{1}{l|}{0.29} &
  \multicolumn{1}{l|}{0.47} &
  \multicolumn{1}{l|}{\textbf{0.18}} &
  \textit{0.91} &
  \multicolumn{1}{l|}{\textbf{0.0757}} &
  \multicolumn{1}{l|}{0.0774} &
  \multicolumn{1}{l|}{\textit{0.0919}} &
  \multicolumn{1}{l|}{0.0852} &
  0.0808 &
  \multicolumn{1}{l|}{8.14} &
  \multicolumn{1}{l|}{8.35} &
  \multicolumn{1}{l|}{\textbf{8.86}} &
  \multicolumn{1}{l|}{8.58} &
  \textit{-0.55} \\
$\text{B}_4$ &
   &
  \multicolumn{1}{l|}{\textbf{0.14}} &
  \multicolumn{1}{l|}{\textbf{0.14}} &
  \multicolumn{1}{l|}{0.19} &
  \multicolumn{1}{l|}{0.16} &
  \textit{1} &
  \multicolumn{1}{l|}{0.1277} &
  \multicolumn{1}{l|}{0.1382} &
  \multicolumn{1}{l|}{0.1288} &
  \multicolumn{1}{l|}{\textbf{0.1143}} &
  \textit{0.1412} &
  \multicolumn{1}{l|}{\textbf{9.21}} &
  \multicolumn{1}{l|}{2.63} &
  \multicolumn{1}{l|}{7.34} &
  \multicolumn{1}{l|}{9.17} &
  \textit{-2.79} \\
$\text{B}_5$ &
   &
  \multicolumn{1}{l|}{0.07} &
  \multicolumn{1}{l|}{0.39} &
  \multicolumn{1}{l|}{\textbf{0.01}} &
  \multicolumn{1}{l|}{0.04} &
  \textit{1} &
  \multicolumn{1}{l|}{0.1467} &
  \multicolumn{1}{l|}{0.1569} &
  \multicolumn{1}{l|}{\textbf{0.1414}} &
  \multicolumn{1}{l|}{0.1427} &
  \textit{0.1812} &
  \multicolumn{1}{l|}{5.19} &
  \multicolumn{1}{l|}{1.77} &
  \multicolumn{1}{l|}{\textbf{5.65}} &
  \multicolumn{1}{l|}{4.24} &
  \textit{-2.54} \\
$\text{B}_6$ &
   &
  \multicolumn{1}{l|}{0.15} &
  \multicolumn{1}{l|}{0.53} &
  \multicolumn{1}{l|}{\textbf{0.07}} &
  \multicolumn{1}{l|}{0.26} &
  \textit{1} &
  \multicolumn{1}{l|}{0.137} &
  \multicolumn{1}{l|}{0.1659} &
  \multicolumn{1}{l|}{\textbf{0.1339}} &
  \multicolumn{1}{l|}{0.1359} &
  \textit{0.1826} &
  \multicolumn{1}{l|}{5.16} &
  \multicolumn{1}{l|}{\textbf{5.78}} &
  \multicolumn{1}{l|}{3.63} &
  \multicolumn{1}{l|}{3.41} &
  \textit{-0.71} \\
$\text{B}_7$ &
   &
  \multicolumn{1}{l|}{\textbf{0.16}} &
  \multicolumn{1}{l|}{0.32} &
  \multicolumn{1}{l|}{\textbf{0.16}} &
  \multicolumn{1}{l|}{0.22} &
  \textit{1} &
  \multicolumn{1}{l|}{\textbf{0.1995}} &
  \multicolumn{1}{l|}{0.2358} &
  \multicolumn{1}{l|}{0.2136} &
  \multicolumn{1}{l|}{0.2255} &
  \textit{0.305} &
  \multicolumn{1}{l|}{-0.65} &
  \multicolumn{1}{l|}{-1.73} &
  \multicolumn{1}{l|}{-0.26} &
  \multicolumn{1}{l|}{\textbf{0.91}} &
  \textit{-7.26} \\
$\text{B}_8$ &
   &
  \multicolumn{1}{l|}{0.06} &
  \multicolumn{1}{l|}{0.42} &
  \multicolumn{1}{l|}{\textbf{0.01}} &
  \multicolumn{1}{l|}{0.24} &
  \textit{0.79} &
  \multicolumn{1}{l|}{0.1634} &
  \multicolumn{1}{l|}{0.1547} &
  \multicolumn{1}{l|}{0.1623} &
  \multicolumn{1}{l|}{\textbf{0.143}} &
  \textit{0.1657} &
  \multicolumn{1}{l|}{\textit{-4.44}} &
  \multicolumn{1}{l|}{-4.41} &
  \multicolumn{1}{l|}{\textbf{2.22}} &
  \multicolumn{1}{l|}{-3.96} &
  -0.71 \\ \hline \hline
$\text{B}_1$ &
  \multirow{8}{*}{Yes} &
  \multicolumn{1}{l|}{0.41} &
  \multicolumn{1}{l|}{0.2} &
  \multicolumn{1}{l|}{\textit{1}} &
  \multicolumn{1}{l|}{\textbf{0.14}} &
  0.83 &
  \multicolumn{1}{l|}{0.1071} &
  \multicolumn{1}{l|}{\textit{0.1139}} &
  \multicolumn{1}{l|}{0.0961} &
  \multicolumn{1}{l|}{\textbf{0.0952}} &
  0.101 &
  \multicolumn{1}{l|}{\textbf{10.09}} &
  \multicolumn{1}{l|}{4.83} &
  \multicolumn{1}{l|}{-9.79} &
  \multicolumn{1}{l|}{9.67} &
  \textit{-14.68} \\
$\text{B}_2$ &
   &
  \multicolumn{1}{l|}{\textit{1}} &
  \multicolumn{1}{l|}{\textbf{0.67}} &
  \multicolumn{1}{l|}{\textit{1}} &
  \multicolumn{1}{l|}{0.8} &
  0.8 &
  \multicolumn{1}{l|}{0.0627} &
  \multicolumn{1}{l|}{0.059} &
  \multicolumn{1}{l|}{\textit{0.0722}} &
  \multicolumn{1}{l|}{\textbf{0.0528}} &
  0.0568 &
  \multicolumn{1}{l|}{6.78} &
  \multicolumn{1}{l|}{2.87} &
  \multicolumn{1}{l|}{-2} &
  \multicolumn{1}{l|}{\textbf{7.73}} &
  \textit{-7.72} \\
$\text{B}_3$ &
   &
  \multicolumn{1}{l|}{0.78} &
  \multicolumn{1}{l|}{\textbf{0.49}} &
  \multicolumn{1}{l|}{\textit{1}} &
  \multicolumn{1}{l|}{0.65} &
  0.85 &
  \multicolumn{1}{l|}{0.0847} &
  \multicolumn{1}{l|}{\textbf{0.0813}} &
  \multicolumn{1}{l|}{\textit{0.1005}} &
  \multicolumn{1}{l|}{0.0953} &
  0.0896 &
  \multicolumn{1}{l|}{3.54} &
  \multicolumn{1}{l|}{3.18} &
  \multicolumn{1}{l|}{-4.35} &
  \multicolumn{1}{l|}{\textbf{7.32}} &
  \textit{-11.82} \\
$\text{B}_4$ &
   &
  \multicolumn{1}{l|}{0.97} &
  \multicolumn{1}{l|}{0.74} &
  \multicolumn{1}{l|}{\textit{1}} &
  \multicolumn{1}{l|}{0.73} &
  \textbf{0.69} &
  \multicolumn{1}{l|}{0.1443} &
  \multicolumn{1}{l|}{0.1296} &
  \multicolumn{1}{l|}{0.1298} &
  \multicolumn{1}{l|}{\textbf{0.1082}} &
  \textit{0.1582} &
  \multicolumn{1}{l|}{0.26} &
  \multicolumn{1}{l|}{0.68} &
  \multicolumn{1}{l|}{-1.19} &
  \multicolumn{1}{l|}{\textbf{3.29}} &
  \textit{-12.43} \\
$\text{B}_5$ &
   &
  \multicolumn{1}{l|}{0.97} &
  \multicolumn{1}{l|}{0.74} &
  \multicolumn{1}{l|}{\textit{1}} &
  \multicolumn{1}{l|}{\textbf{0.7}} &
  \textit{1} &
  \multicolumn{1}{l|}{0.158} &
  \multicolumn{1}{l|}{0.16} &
  \multicolumn{1}{l|}{\textbf{0.1522}} &
  \multicolumn{1}{l|}{0.1552} &
  \textit{0.1982} &
  \multicolumn{1}{l|}{\textbf{0.93}} &
  \multicolumn{1}{l|}{-1.69} &
  \multicolumn{1}{l|}{-6.41} &
  \multicolumn{1}{l|}{0.07} &
  \textit{-13.54} \\
$\text{B}_6$ &
   &
  \multicolumn{1}{l|}{0.8} &
  \multicolumn{1}{l|}{0.71} &
  \multicolumn{1}{l|}{\textit{1}} &
  \multicolumn{1}{l|}{0.64} &
  \textbf{0.26} &
  \multicolumn{1}{l|}{0.1606} &
  \multicolumn{1}{l|}{0.1673} &
  \multicolumn{1}{l|}{\textbf{0.145}} &
  \multicolumn{1}{l|}{0.1525} &
  \textit{0.1836} &
  \multicolumn{1}{l|}{0.34} &
  \multicolumn{1}{l|}{-2.08} &
  \multicolumn{1}{l|}{-3.27} &
  \multicolumn{1}{l|}{\textbf{3.21}} &
  \textit{-4.23} \\
$\text{B}_7$ &
   &
  \multicolumn{1}{l|}{\textit{1}} &
  \multicolumn{1}{l|}{\textbf{0.14}} &
  \multicolumn{1}{l|}{\textit{1}} &
  \multicolumn{1}{l|}{0.56} &
  0.7 &
  \multicolumn{1}{l|}{0.2252} &
  \multicolumn{1}{l|}{\textbf{0.2116}} &
  \multicolumn{1}{l|}{0.2135} &
  \multicolumn{1}{l|}{0.2137} &
  \textit{0.3168} &
  \multicolumn{1}{l|}{\textbf{-2.33}} &
  \multicolumn{1}{l|}{-3.48} &
  \multicolumn{1}{l|}{-3.89} &
  \multicolumn{1}{l|}{-4.42} &
  \textit{-12.38} \\
$\text{B}_8$ &
   &
  \multicolumn{1}{l|}{0.93} &
  \multicolumn{1}{l|}{\textbf{0.61}} &
  \multicolumn{1}{l|}{\textit{1}} &
  \multicolumn{1}{l|}{0.63} &
  \textit{1} &
  \multicolumn{1}{l|}{0.1928} &
  \multicolumn{1}{l|}{0.1871} &
  \multicolumn{1}{l|}{\textit{0.2205}} &
  \multicolumn{1}{l|}{\textbf{0.1628}} &
  0.1822 &
  \multicolumn{1}{l|}{\textit{-7.62}} &
  \multicolumn{1}{l|}{-3.92} &
  \multicolumn{1}{l|}{\textbf{-1.25}} &
  \multicolumn{1}{l|}{-5.22} &
  -0.77 \\ \hline
\end{tabular}
}
\label{tab:benchmark_results}
\end{table*}

\section{Benchmark-Based Evaluation Results}
In this section, we test the ability of RIS $\text{S}_1$-$\text{S}_5$ in reproducing $\text{B}_1$-$\text{B}_8$. The reference RIS S$_0$ is also used to reproduce the same benchmarking patterns. Each RIS (including S$_0$) has 40x40 (1600) unit-cells. A commercially available full-wave EM solver (CST) is used for unit-cell simulations to obtain their radiation responses. We use the lumped element model (s-parameters) of PIN diodes from Skyworks SMP1340-040LF. We use the unit-cell data from CST and array factor approximation to simulate the RIS radiation patterns under point/planewave source assumptions. A genetic algorithm (GA) implemented in MATLAB determines the optimal configuration (diode ON/OFF states) of unit-cells on each RIS to generate benchmarking patterns. The finite-sized RISs with discrete phase controls can fail to generate the desired radiation patterns exactly. Therefore, in our simulations, we allow GA to stop after 350 generations because we observe that there is no significant improvement even if we allow the algorithm to run longer. When the GA stops, we take the best match. We use the patterns produced by S$_0$ as reference radiation patterns to compute DE, NMSE, and SLR metrics.

\subsection{Case 1: Point Source}
The qualitative and quantitative results under point source assumptions are presented in Figure~\ref{fig:benchmark_plots} and Table~\ref{tab:benchmark_results}. The impact of unit-cell grouping to reduce the control circuit complexity is given in Table~\ref{tab:benchmark_results}.

\subsubsection{Performance Without Unit-Cell Grouping}
In Figure~\ref{fig:benchmark_plots}, we present the visual results that show the quality of the reproduced radiation patterns by each RIS when we individually control the unit-cells. The performance of S$_0$ is also plotted. 
\\ \textbf{RIS - $\text{S}_1$}: The visual quality of $\text{B}_1$-$\text{B}_6$ looks excellent, while that of $\text{B}_7$ and $\text{B}_8$ is poor. There are few undesired main lobes in $\text{B}_6$, $\text{B}_7$ and $\text{B}_8$ as well as several high power side lobes, especially in $\text{B}_8$. However, the value of DE for $\text{B}_8$ is very small. The observations about the excellent quality of radiation patterns are confirmed by corresponding values of DE, NMSE and SLR in each case, which is either the best or very close to the best values.  
\\ \textbf{RIS - $\text{S}_2$}: The visual quality of radiation patterns produced by $\text{S}_2$ looks somewhat poorer to the ones produced by $\text{S}_1$. From Table~\ref{tab:benchmark_results}, we find out that DE (except for $\text{B}_1$) and SLR (except for $\text{B}_6$, $\text{B}_8$) values are poor compared to $\text{S}_1$ on almost all the benchmarks. From Table~\ref{tab:sec3table}, the control circuit complexity and power requirements of both $\text{S}_1$ and $\text{S}_2$ are the same. Therefore, among optimized 1-bit RISs, $\text{S}_1$ comes out as a better choice than $\text{S}_2$ on nearly all the performance metrics under point source assumption without unit-cell grouping.  
\\ \textbf{RIS - $\text{S}_3$}: The visual quality of $\text{B}_1$-$\text{B}_6$ produced by $\text{S}_3$ is similar to $\text{S}_1$ and better in more complex benchmarks. Moreover, the quality of $\text{B}_6$-$\text{B}_8$ looks significantly better than $\text{S}_1$ and $\text{S}_2$. The quantitative metrics are also better for several benchmarking patterns. From Table~\ref{tab:sec3table}, the control circuit complexity of $\text{S}_3$ is double while its function switching rate is half than both $\text{S}_1$ and $\text{S}_2$. Since $\text{S}_3$ is designed to operate in S-band, its per unit area power requirements are significantly less than both $\text{S}_1$ and $\text{S}_2$. However, the total power requirements of $\text{S}_3$ are the highest among all the RIS because each unit-cell has 5 PIN diodes. 
\\ \textbf{RIS - $\text{S}_4$}: The visual quality of $\text{B}_2$-$\text{B}_6$ produced by $\text{S}_4$ is similar to $\text{S}_1$, $\text{S}_2$ and $\text{S}_3$. The quality of $\text{B}_1$-$\text{B}_8$ looks worse than $\text{S}_3$ and $\text{S}_1$ but better than $\text{S}_2$. We notice that DE is not significantly poor, but the performance is effectively degraded due to higher side lobes and scattering. The control circuit complexity and power requirements of $\text{S}_4$ are double than both $\text{S}_1$ and $\text{S}_2$. However, this RIS uses only 2 PIN diodes per unit-cell, significantly reducing its power requirements compared to $\text{S}_3$.  
\\ \textbf{RIS - $\text{S}_5$}: This surface is made from a 1-bit unoptimized unit-cell. The visual quality on all the benchmarks is worse than all other RIS. However, on $\text{B}_2$, the main lobes are visible; one is along the desired direction with two high-power side lobes. The visual quality on $\text{B}_1$-$\text{B}_8$ is inferior because there is no correlation between the target pattern and the pattern produced by $\text{S}_5$, i.e., $\text{S}_5$ is incapable of beam steering except the simple case of very small angle. The control circuit complexity and power requirements of $\text{S}_5$ are similar to $\text{S}_1$ and $\text{S}_2$.

In terms of reproducing $\text{B}_1$-$\text{B}_8$ under point source assumptions without unit-cell grouping, the overall performance of $\text{S}_3$ is the best, followed by $\text{S}_1$, $\text{S}_4$, $\text{S}_2$ and $\text{S}_5$. Considering control circuit complexity and power requirements along with the DE, NMSE, and SLR, $\text{S}_1$ seems a better choice. However, in application scenarios where $\text{B}_6$-$\text{B}_8$ are mostly required, $\text{S}_3$ should be preferred despite its complexity and power requirements. Finally, if the application scenario only demands small angle beam steering (such as $\text{B}_2$), we may also use $\text{S}_5$ or its slightly more optimized version.  

\subsubsection{Performance With Unit-Cell Grouping}
We now consider RIS $\text{S}_1$-$\text{S}_5$ assuming the unit-cells are grouped into pairs. This way, the RIS's complexity, and sensing overhead is halved. However, the maximum power requirements remain unchanged. The impact on the resulting RISs' ability to produce the benchmarking patterns is quantified in the lower half of Table~\ref{tab:benchmark_results}. We can see that all metrics show poorer performance for all the designs. However, the extent to which unit-cell grouping impacts the performance of each design is different. 

When unit-cells are controlled in pairs, the best performance is shown by $\text{S}_2$ followed by $\text{S}_4$, $\text{S}_1$, $\text{S}_5$ and $\text{S}_3$. Please note that the DE value of 1 means that the surface has failed to produce the main beams in the intended directions. Poor performing surfaces without grouping ($\text{S}_2$, $\text{S}_4$, $\text{S}_5$) are observed to be least affected by unit-cell grouping. We should expect small gains in their performance even if we further increase the RIS size by increasing the individually controlled unit-cells. On the other hand, the gains in the performance of $\text{S}_1$ and $\text{S}_3$ should become significantly higher with the corresponding increase in the RIS size. Meanwhile, the choice of beam steering angles and the variation in beam power levels can significantly raise the required performance criteria for RIS designers, as we witness in the case of B$_8$.

\begin{figure*}
    \centering
    \includegraphics[width=0.98\textwidth]{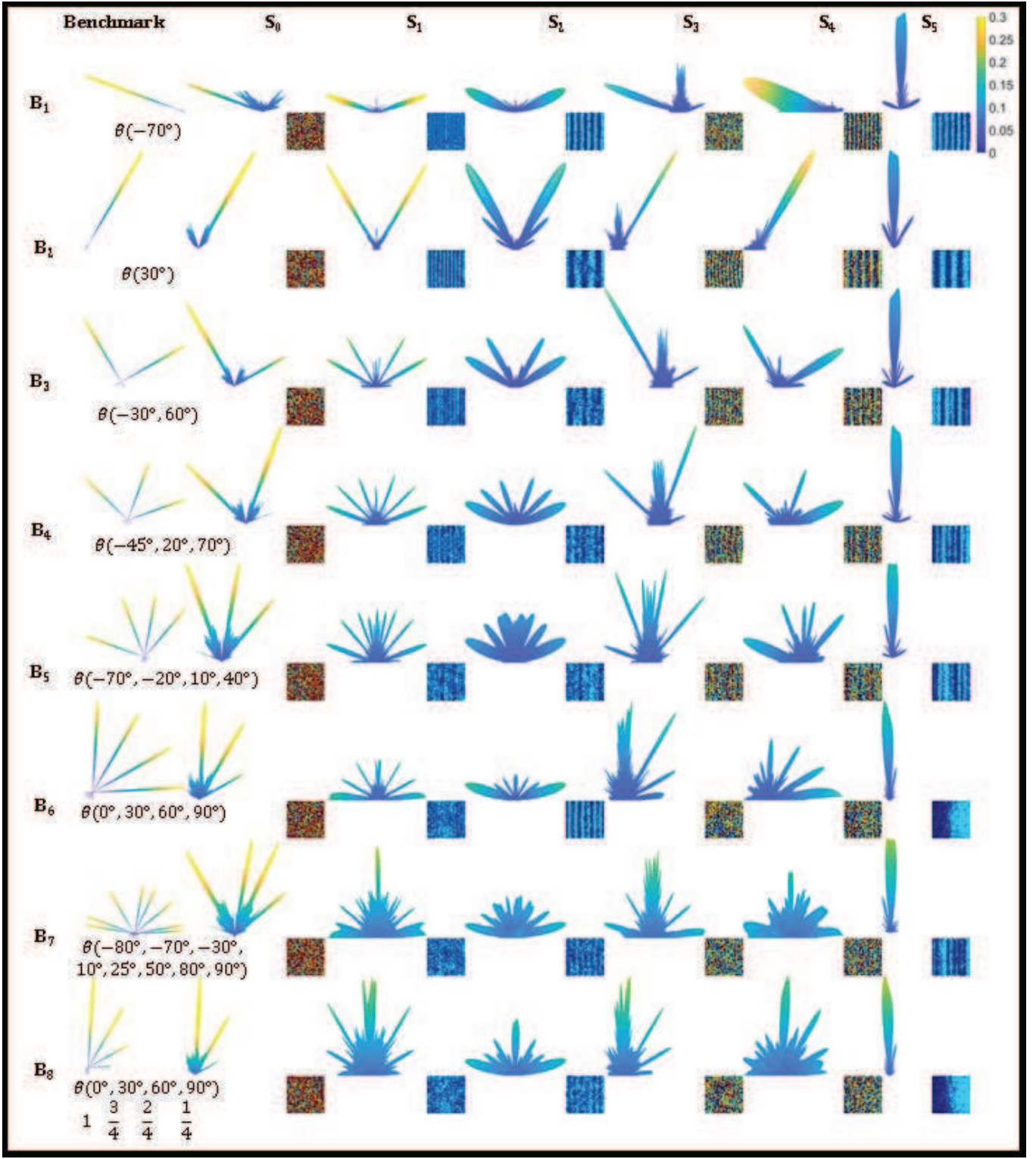}
    \caption{Benchmarking radiation patterns $\text{B}_1$-$\text{B}_8$ and the radiation patterns obtained by RIS $\text{S}_1$-$\text{S}_5$ under planewave source assumptions (case 2) without unit-cell grouping. The results for the reference RIS S$_0$ are also shown. The RIS configuration matrix ($40\times40$) is also reported as a colored image (unit-cell state 1=Blue, 2=Cyan, 3=Yellow, 4=Red). The elevation identifies Beam directions ($\theta$, ranges between 0$^\circ$ and 90$^\circ$) and azimuth($\phi$, only takes two values 0$^\circ$ and 180$^\circ$) angles. Instead of repeating the two $\phi$ values, $\theta$ and $-\theta$ are used ($-\theta = \theta_{\phi=180^\circ}$).}
    \label{fig:pbenchmark_plots}
\end{figure*}

\begin{table*}[t]
\centering
\caption{Quantitative results of five RIS S$_{1}-\text{S}_5$, with and without unit-cell grouping under planewave source assumptions with the observer in the far-field (case 2). Best values (smallest in case of DE and NMSE and largest in case of SLR) are highlighted in \textbf{bold} font, worst values (largest in case of DE and NMSE and smallest in case of SLR) are highlighted in \textit{italic} font.}
\resizebox{\textwidth}{!}{
\renewcommand{\arraystretch}{1.2}
\begin{tabular}{|c|c|lllll||lllll||lllll|}
\hline
ID &
  Grouping &
  \multicolumn{5}{c|}{Relative Directivity Error (DE)} &
  \multicolumn{5}{c|}{Relative Normalized Mean Squared Error (NMSE)} &
  \multicolumn{5}{c|}{Side Lobe Ratio (SLR) - dB} \\ \hline
 &
   &
  \multicolumn{1}{c|}{$S_1$} &
  \multicolumn{1}{c|}{$S_2$} &
  \multicolumn{1}{c|}{$S_3$} &
  \multicolumn{1}{c|}{$S_4$} &
  \multicolumn{1}{c|}{$S_5$} &
  \multicolumn{1}{c|}{$S_1$} &
  \multicolumn{1}{c|}{$S_2$} &
  \multicolumn{1}{c|}{$S_3$} &
  \multicolumn{1}{c|}{$S_4$} &
  \multicolumn{1}{c|}{$S_5$} &
  \multicolumn{1}{c|}{$S_1$} &
  \multicolumn{1}{c|}{$S_2$} &
  \multicolumn{1}{c|}{$S_3$} &
  \multicolumn{1}{c|}{$S_4$} &
  \multicolumn{1}{c|}{$S_5$} \\ \hline
$\text{B}_1$ &
  \multirow{8}{*}{No} &
  \multicolumn{1}{l|}{\textbf{-0.13}} &
  \multicolumn{1}{l|}{0.73} &
  \multicolumn{1}{l|}{\textit{1}} &
  \multicolumn{1}{l|}{0.15} &
  \textit{1} &
  \multicolumn{1}{l|}{\textit{0.0166}} &
  \multicolumn{1}{l|}{0.0138} &
  \multicolumn{1}{l|}{0.014} &
  \multicolumn{1}{l|}{\textbf{0.0102}} &
  0.0158 &
  \multicolumn{1}{l|}{0.04} &
  \multicolumn{1}{l|}{0.17} &
  \multicolumn{1}{l|}{-8.51} &
  \multicolumn{1}{l|}{\textbf{14.14}} &
  \textit{-14.03} \\
$\text{B}_2$ &
   &
  \multicolumn{1}{l|}{\textbf{0.12}} &
  \multicolumn{1}{l|}{0.93} &
  \multicolumn{1}{l|}{0.99} &
  \multicolumn{1}{l|}{0.55} &
  \textit{1} &
  \multicolumn{1}{l|}{\textit{0.01}} &
  \multicolumn{1}{l|}{0.0092} &
  \multicolumn{1}{l|}{0.0082} &
  \multicolumn{1}{l|}{\textbf{0.0049}} &
  0.0097 &
  \multicolumn{1}{l|}{-0.03} &
  \multicolumn{1}{l|}{0.05} &
  \multicolumn{1}{l|}{-3.28} &
  \multicolumn{1}{l|}{\textbf{10.2}} &
  \textit{-8.82} \\
$\text{B}_3$ &
   &
  \multicolumn{1}{l|}{\textbf{0.61}} &
  \multicolumn{1}{l|}{0.89} &
  \multicolumn{1}{l|}{0.98} &
  \multicolumn{1}{l|}{0.8} &
  \textit{1} &
  \multicolumn{1}{l|}{0.0119} &
  \multicolumn{1}{l|}{0.0112} &
  \multicolumn{1}{l|}{0.0113} &
  \multicolumn{1}{l|}{\textbf{0.0074}} &
  \textit{0.0129} &
  \multicolumn{1}{l|}{-0.09} &
  \multicolumn{1}{l|}{0.02} &
  \multicolumn{1}{l|}{-8.34} &
  \multicolumn{1}{l|}{\textbf{6}} &
  \textit{-12.97} \\
$\text{B}_4$ &
   &
  \multicolumn{1}{l|}{0.78} &
  \multicolumn{1}{l|}{0.91} &
  \multicolumn{1}{l|}{\textit{0.99}} &
  \multicolumn{1}{l|}{\textbf{0.76}} &
  \textit{0.99} &
  \multicolumn{1}{l|}{0.0117} &
  \multicolumn{1}{l|}{0.0116} &
  \multicolumn{1}{l|}{0.0121} &
  \multicolumn{1}{l|}{\textbf{0.0083}} &
  \textit{0.0141} &
  \multicolumn{1}{l|}{-0.17} &
  \multicolumn{1}{l|}{-0.49} &
  \multicolumn{1}{l|}{-9.84} &
  \multicolumn{1}{l|}{\textbf{5.7}} &
  \textit{-16.3} \\
$\text{B}_5$ &
   &
  \multicolumn{1}{l|}{\textbf{0.86}} &
  \multicolumn{1}{l|}{0.92} &
  \multicolumn{1}{l|}{\textit{0.93}} &
  \multicolumn{1}{l|}{0.88} &
  0.87 &
  \multicolumn{1}{l|}{0.0139} &
  \multicolumn{1}{l|}{0.0142} &
  \multicolumn{1}{l|}{0.0158} &
  \multicolumn{1}{l|}{\textbf{0.0112}} &
  \textit{0.0184} &
  \multicolumn{1}{l|}{-0.14} &
  \multicolumn{1}{l|}{-3.38} &
  \multicolumn{1}{l|}{-11.79} &
  \multicolumn{1}{l|}{\textbf{0.98}} &
  \textit{-15.62} \\
$\text{B}_6$ &
   &
  \multicolumn{1}{l|}{0.6} &
  \multicolumn{1}{l|}{\textbf{0.52}} &
  \multicolumn{1}{l|}{0.84} &
  \multicolumn{1}{l|}{0.75} &
  \textit{0.88} &
  \multicolumn{1}{l|}{0.0141} &
  \multicolumn{1}{l|}{\textit{0.0171}} &
  \multicolumn{1}{l|}{0.0119} &
  \multicolumn{1}{l|}{\textbf{0.0079}} &
  0.0154 &
  \multicolumn{1}{l|}{-1.58} &
  \multicolumn{1}{l|}{-3.19} &
  \multicolumn{1}{l|}{\textit{-8.15}} &
  \multicolumn{1}{l|}{\textbf{2.39}} &
  -6.49 \\
$\text{B}_7$ &
   &
  \multicolumn{1}{l|}{0.97} &
  \multicolumn{1}{l|}{0.97} &
  \multicolumn{1}{l|}{0.92} &
  \multicolumn{1}{l|}{\textit{0.98}} &
  \textbf{-0.05} &
  \multicolumn{1}{l|}{0.0185} &
  \multicolumn{1}{l|}{0.0184} &
  \multicolumn{1}{l|}{0.0237} &
  \multicolumn{1}{l|}{\textbf{0.018}} &
  \textit{0.0259} &
  \multicolumn{1}{l|}{-5.68} &
  \multicolumn{1}{l|}{\textbf{-3.29}} &
  \multicolumn{1}{l|}{-16.15} &
  \multicolumn{1}{l|}{-4.74} &
  \textit{-20.78} \\
$\text{B}_8$ &
   &
  \multicolumn{1}{l|}{\textit{1}} &
  \multicolumn{1}{l|}{\textit{1}} &
  \multicolumn{1}{l|}{\textit{1}} &
  \multicolumn{1}{l|}{\textbf{0.99}} &
  \textit{1} &
  \multicolumn{1}{l|}{0.0131} &
  \multicolumn{1}{l|}{0.0131} &
  \multicolumn{1}{l|}{0.0148} &
  \multicolumn{1}{l|}{\textbf{0.012}} &
  \textit{0.0171} &
  \multicolumn{1}{l|}{-4.72} &
  \multicolumn{1}{l|}{3.58} &
  \multicolumn{1}{l|}{-1.25} &
  \multicolumn{1}{l|}{\textit{-6.15}} &
  \textbf{5.04} \\ \hline \hline
$\text{B}_1$ &
  \multirow{8}{*}{Yes} &
  \multicolumn{1}{l|}{0.16} &
  \multicolumn{1}{l|}{0.74} &
  \multicolumn{1}{l|}{\textit{1}} &
  \multicolumn{1}{l|}{\textbf{0.14}} &
  0.98 &
  \multicolumn{1}{l|}{0.1988} &
  \multicolumn{1}{l|}{\textbf{0.1553}} &
  \multicolumn{1}{l|}{\textit{0.3117}} &
  \multicolumn{1}{l|}{0.1762} &
  0.2235 &
  \multicolumn{1}{l|}{-0.19} &
  \multicolumn{1}{l|}{-0.264} &
  \multicolumn{1}{l|}{-1.958} &
  \multicolumn{1}{l|}{\textbf{2.704}} &
  \textit{-3.088} \\
$\text{B}_2$ &
   &
  \multicolumn{1}{l|}{\textbf{0.13}} &
  \multicolumn{1}{l|}{\textit{1}} &
  \multicolumn{1}{l|}{\textit{1}} &
  \multicolumn{1}{l|}{0.52} &
  \textit{1} &
  \multicolumn{1}{l|}{\textit{0.2335}} &
  \multicolumn{1}{l|}{0.1829} &
  \multicolumn{1}{l|}{0.1096} &
  \multicolumn{1}{l|}{\textbf{0.0317}} &
  0.1279 &
  \multicolumn{1}{l|}{-0.172} &
  \multicolumn{1}{l|}{-0.276} &
  \multicolumn{1}{l|}{-0.94} &
  \multicolumn{1}{l|}{\textbf{1.936}} &
  \textit{-1.918} \\
$\text{B}_3$ &
   &
  \multicolumn{1}{l|}{\textbf{0.82}} &
  \multicolumn{1}{l|}{\textit{1}} &
  \multicolumn{1}{l|}{\textit{1}} &
  \multicolumn{1}{l|}{\textit{1}} &
  \textit{1} &
  \multicolumn{1}{l|}{0.1773} &
  \multicolumn{1}{l|}{\textit{0.2782}} &
  \multicolumn{1}{l|}{0.2049} &
  \multicolumn{1}{l|}{\textbf{0.0996}} &
  0.2776 &
  \multicolumn{1}{l|}{-0.308} &
  \multicolumn{1}{l|}{-0.162} &
  \multicolumn{1}{l|}{-1.926} &
  \multicolumn{1}{l|}{\textbf{1.098}} &
  \textit{-2.77} \\
$\text{B}_4$ &
   &
  \multicolumn{1}{l|}{\textit{1}} &
  \multicolumn{1}{l|}{0.95} &
  \multicolumn{1}{l|}{\textit{1}} &
  \multicolumn{1}{l|}{\textbf{0.85}} &
  \textit{1} &
  \multicolumn{1}{l|}{0.2178} &
  \multicolumn{1}{l|}{\textbf{0.0701}} &
  \multicolumn{1}{l|}{0.0887} &
  \multicolumn{1}{l|}{0.1299} &
  \textit{0.3023} &
  \multicolumn{1}{l|}{-0.15} &
  \multicolumn{1}{l|}{-0.266} &
  \multicolumn{1}{l|}{-2.154} &
  \multicolumn{1}{l|}{\textbf{0.918}} &
  \textit{-3.474} \\
$\text{B}_5$ &
   &
  \multicolumn{1}{l|}{\textit{1}} &
  \multicolumn{1}{l|}{\textit{1}} &
  \multicolumn{1}{l|}{0.99} &
  \multicolumn{1}{l|}{\textit{1}} &
  \textbf{0.81} &
  \multicolumn{1}{l|}{0.1942} &
  \multicolumn{1}{l|}{\textit{0.2579}} &
  \multicolumn{1}{l|}{0.2359} &
  \multicolumn{1}{l|}{0.1888} &
  \textbf{0.1287} &
  \multicolumn{1}{l|}{-0.292} &
  \multicolumn{1}{l|}{-0.848} &
  \multicolumn{1}{l|}{-2.578} &
  \multicolumn{1}{l|}{\textbf{-0.018}} &
  \textit{-3.288} \\
$\text{B}_6$ &
   &
  \multicolumn{1}{l|}{0.72} &
  \multicolumn{1}{l|}{\textbf{0.69}} &
  \multicolumn{1}{l|}{\textit{1}} &
  \multicolumn{1}{l|}{0.87} &
  0.96 &
  \multicolumn{1}{l|}{0.2251} &
  \multicolumn{1}{l|}{\textit{0.3974}} &
  \multicolumn{1}{l|}{0.1984} &
  \multicolumn{1}{l|}{\textbf{0.0493}} &
  0.129 &
  \multicolumn{1}{l|}{-0.524} &
  \multicolumn{1}{l|}{-0.738} &
  \multicolumn{1}{l|}{\textit{-1.734}} &
  \multicolumn{1}{l|}{\textbf{0.232}} &
  -1.404 \\
$\text{B}_7$ &
   &
  \multicolumn{1}{l|}{0.94} &
  \multicolumn{1}{l|}{\textit{1}} &
  \multicolumn{1}{l|}{\textit{1}} &
  \multicolumn{1}{l|}{0.94} &
  \textbf{0.09} &
  \multicolumn{1}{l|}{0.2183} &
  \multicolumn{1}{l|}{\textit{0.3955}} &
  \multicolumn{1}{l|}{\textbf{0.1911}} &
  \multicolumn{1}{l|}{0.2659} &
  0.3232 &
  \multicolumn{1}{l|}{-1.328} &
  \multicolumn{1}{l|}{\textbf{-0.926}} &
  \multicolumn{1}{l|}{-3.334} &
  \multicolumn{1}{l|}{-1.164} &
  \textit{-4.384} \\
$\text{B}_8$ &
   &
  \multicolumn{1}{l|}{\textit{1}} &
  \multicolumn{1}{l|}{\textbf{0.98}} &
  \multicolumn{1}{l|}{\textit{1}} &
  \multicolumn{1}{l|}{\textit{1}} &
  \textit{1} &
  \multicolumn{1}{l|}{\textbf{0.0649}} &
  \multicolumn{1}{l|}{0.1213} &
  \multicolumn{1}{l|}{0.2442} &
  \multicolumn{1}{l|}{0.1574} &
  \textit{0.3597} &
  \multicolumn{1}{l|}{-1.236} &
  \multicolumn{1}{l|}{0.424} &
  \multicolumn{1}{l|}{-0.368} &
  \multicolumn{1}{l|}{\textit{-1.43}} &
  \textbf{0.88} \\ \hline
\end{tabular}
}
\label{tab:benchmark_results_planewave}
\end{table*}

\subsection{Case 2: Planewave Source}
The qualitative and quantitative results under planewave source assumptions are presented in Figure~\ref{fig:pbenchmark_plots} and Table~\ref{tab:benchmark_results_planewave}. In this case, the number of unit-cell control states greatly impacts the performance. 

\subsubsection{Performance Without Unit-Cell Grouping}
The qualitative results without unit-cell grouping are shown in Figure~\ref{fig:pbenchmark_plots}, while the corresponding quantitative results are presented in the upper half of Table~\ref{tab:benchmark_results_planewave}.\\
\noindent \textbf{RIS - $\text{S}_1$}: We can observe that the RIS can reproduce the main lobes, but there are non-intended side lobes in every radiation pattern produced by $\text{S}_1$. These side lobes are called `quantization lobes.' They generally appear in RISs made from 1-bit unit-cells due to more pronounced phase rounding quantization effects~\cite{kashyap2020mitigating,smith1983comparison}. The RIS can reproduce the main lobes; therefore, the DE values are very good in four out of eight benchmarking patterns ($\text{B}_1$-$\text{B}_3$, $\text{B}_5$), but NMSE is very bad compared to the best NMSE obtained for RIS $\text{S}_4$ in the same tests. The side lobe level of $\text{S}_1$ is close to 0 or negative in some cases, showing that the scattering level is close to the main lobes. 
\\ \textbf{RIS - $\text{S}_2$}: This 1-bit RIS also produces quantization lobes. We can observe that main lobes are wider compared to $\text{S}_1$. On quantitative metrics, except $\text{B}_6$, the performance of $\text{S}_2$ is generally poor and NMSE is very close to $\text{S}_1$ in all the tests. The side lobes are close to the main lobes or even greater in magnitude than the main lobes making the SLR values close to 0 or negative. 
\\ \textbf{RIS - $\text{S}_3$}: There are no quantization lobes in the radiation patterns produced by $\text{S}_3$. The main lobes also look narrow and focused. However, the RIS could not form the main lobes in the intended directions. The starting and ending angles of the main lobes in the benchmarking and the achieved radiation patterns are very close to each other but there is no overlap due to narrow beams formed by $\text{S}_3$. Thus, the DE and SLR values are extremely poor because although beam steering is prominent but it is not in the intended directions. On the other hand, NMSE values are relatively good and second best in two cases ($\text{B}_2$ and $\text{B}_3$).  
\\ \textbf{RIS - $\text{S}_4$}: The visual quality of the radiation patterns produced by $\text{S}_4$ also looks good, and the quantization lobes are also missing. The overall performance of $\text{S}_4$ is significantly better in all the benchmark tests. The RIS can produce the main lobes with low side lobe levels. Moreover, the main lobes are wide as compared to $\text{S}_3$ and the starting and ending angles of the main lobes in the benchmarking and the achieved radiation patterns overlap. Therefore, the DE values are better than $\text{S}_3$ but relatively worse than $\text{S}_1$ on multiple benchmarks under planewave source assumptions. However, the values of NMSE and SLR in all benchmark tests are significantly better (often the best) than $\text{S}_1$ and $\text{S}_3$. 
\\ \textbf{RIS - $\text{S}_5$}: This RIS consists of an un-optimized 1-bit unit-cell, and it is evident that the surface has entirely failed in almost all the benchmarking patterns. The visual quality is the worst because the RIS lacks the beam steering capability under planewave source assumption. The quantitative values of $\text{S}_5$ are also the worst on all the benchmarking patterns. In $\text{B}_7$, the DE is close to 0, but NMSE and SLR are still the worst. The better value of DE occurs because $\text{B}_7$ has eight beams, and high scattering can at least satisfy some intended directions. Still, poor performance becomes evident when the error of unintended directions is considered through NMSE and SLR values. The same is true for $\text{B}_8$ where SLR is good, but the other two metrics are the worst. 

\subsubsection{Performance With Unit-Cell Grouping} When we group the unit-cells in pairs, the performance of all the RISs under planewave source assumptions further decreases. As shown in the bottom half of Table~\ref{tab:benchmark_results_planewave} RISs $\text{S}_1$ and $\text{S}_2$ consisting of 1-bit unit-cells shows degraded performance on all the metrics as compared to RISs $\text{S}_3$ and $\text{S}_4$ consisting of 2-bit unit-cells. The performance of RIS $\text{S}_5$ is still the worst. The distinction between 1-bit and 2-bit unit-cell and the effect of unit-cell radiation response are more prominent in dictating RIS beam steering capability under planewave source assumptions. Optimizing unit-cell in terms of producing maximally apart phase shifts in different control states also becomes crucial in obtaining good performance. 

\subsection{Discussion and Comparison}
\begin{figure}
    \centering
    \includegraphics[width=\columnwidth]{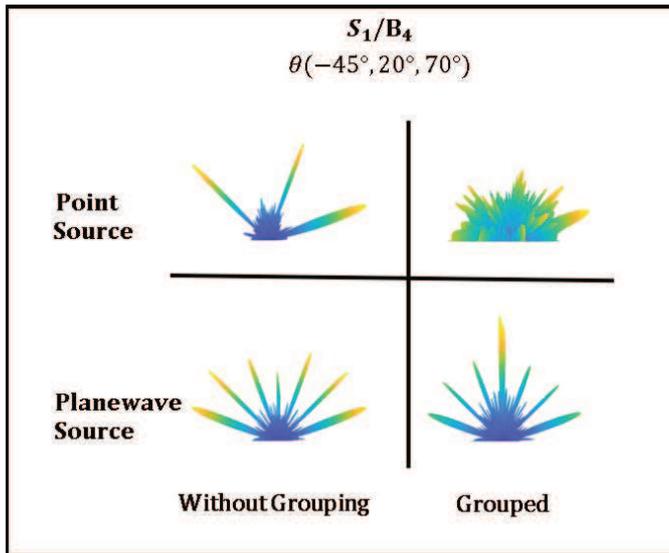}
    \caption{Radiation patterns generated by $\text{S}_1$ (1-bit unit-cell) while reproducing the benchmarking pattern $\text{B}_4$ under point/planewave source assumptions and without/with unit-cell grouping in pairs.}
    \label{fig:s1_b4_plots}
\end{figure}

In Figure \ref{fig:s1_b4_plots}, we present the visual quality of the radiation pattern produced by $\text{S}_1$ (1-bit unit-cell) while reproducing the benchmarking pattern $\text{B}_4$ under point/planewave source assumptions and without/with unit-cell grouping in pairs. We can see that we obtain the best result under the point source assumption without grouping, which means that all the 1600 1-bit unit-cells on the RIS have a separate control. On the other hand, when we control the unit-cells in pairs, under point source assumptions, the ability of the surface to reproduce the given pattern is severely degraded with a significant scattering of power in all directions. Under planewave source assumptions and without unit-cell grouping, the quantization lobes are visible along with all the desired main lobes. When the unit-cells are grouped in pairs to reduce the control circuit overhead, the performance degrades under planewave source assumption.   

\begin{figure}
    \centering
    \includegraphics[width=\columnwidth]{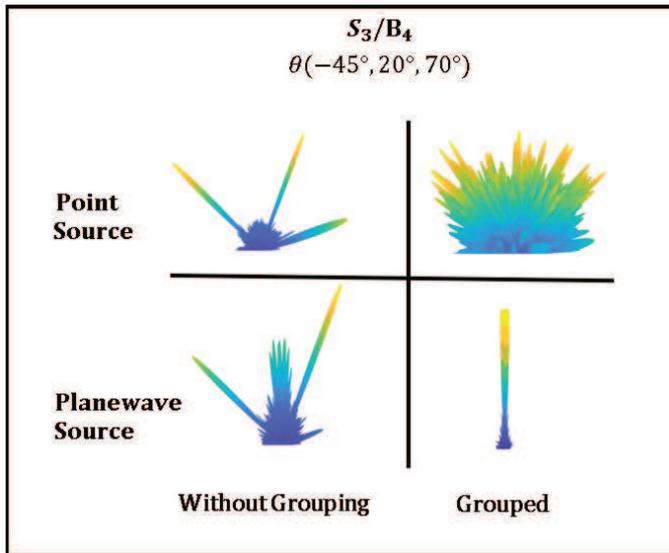}
    \caption{Radiation patterns generated by $\text{S}_3$ (2-bit unit-cell) while reproducing the benchmarking pattern $\text{B}_4$ under point/planewave source assumptions and without/with unit-cell grouping in pairs.}
    \label{fig:s3_b4_plots}
\end{figure}

In Figure \ref{fig:s3_b4_plots}, we present the visual quality of the radiation pattern produced by $\text{S}_3$ (2-bit unit-cell) while reproducing the benchmarking pattern $\text{B}_4$ under point/planewave source assumptions and without/with unit-cell grouping in pairs. Under the point source assumption, without unit-cell grouping, the three beams are visible. In this case, the scattering level in the unintended directions is relatively less compared to the radiation pattern generated by $\text{S}_1$. When we group the unit-cells in pairs, the performance significantly degrades under the point source assumption, and it becomes equally bad, as we observe in $\text{S}_1$. Under planewave source assumptions and without unit-cell grouping, the performance of $\text{S}_3$ is comparable to the point source case with two out of three main lobes visible. Unlike $\text{S}_1$, there are no quantization lobes. Finally, with grouping and planewave source assumptions, $\text{S}_3$ loses its beam steering capability and cannot form any of the major lobes. 

The interplay of unit-cell design, grouping for control circuit complexity reduction, the number of diode control states on each unit-cell, the unit-cell radiation response, the power requirements of the RIS control circuit, and the radiation pattern generation capabilities of RIS under point/planewave source assumptions is evident from these results. Our benchmarking framework, metrics, and these observations would help the research community make appropriate design choices and tradeoffs to achieve the desired far-field beamforming performance in different smart city application scenarios.

\section{Conclusion}
We conducted a detailed analysis of the performance aspects of RISs composed of various unit-cell designs. We developed a benchmarking framework that included radiation patterns commonly required in smart city environments and presented performance metrics to quantify RIS radiation pattern generation capabilities relative to the benchmarking patterns, control circuit complexity, and power requirements for point/planewave source assumptions. Using the framework, we tested and compared five different RISs ($\text{S}_1$-$\text{S}_5$) made up from five different unit-cell designs. The unit-cells on these surfaces were first individually controlled and later in groups. The proposed framework can be handy for choosing unit-cells from existing designs in various application settings and source type assumptions. We can also use the framework to determine the usefulness of future designs and their most viable applications. With the help of our proposed framework, we can also determine the optimal group size for the best tradeoff that favors ease of implementation and reduces the channel sensing overhead. A better understanding of the unit-cell- and surface-level RIS design interplay could lead to better strategies for system-level performance enhancement in RIS-assisted communications.

\bibliographystyle{IEEEtran}   
\bibliography{references} 

\end{document}